\documentclass[journal=nalefd,manuscript=letter]{achemso}

\usepackage[version=3]{mhchem} 
\usepackage{amssymb}
\usepackage{siunitx}

\DeclareSIUnit\angstrom{\text{Å}}

\author{Sani Y. Harouna-Mayer}
\affiliation{Institute for Nanostructure and Solid-State Physics, Center for Hybrid Nanostructures, University of Hamburg, Hamburg 22761, Germany}
\altaffiliation{The Hamburg Center for Ultrafast Imaging, Hamburg 22761, Germany}
\author{Lars Klemeyer}
\affiliation{Institute for Nanostructure and Solid-State Physics, Center for Hybrid Nanostructures, University of Hamburg, Hamburg 22761, Germany}
\altaffiliation{The Hamburg Center for Ultrafast Imaging, Hamburg 22761, Germany}
\author{Cecilia A. Zito}
\affiliation{Institute for Nanostructure and Solid-State Physics, Center for Hybrid Nanostructures, University of Hamburg, Hamburg 22761, Germany}
\altaffiliation{The Hamburg Center for Ultrafast Imaging, Hamburg 22761, Germany}
\author{Johan Bielecki}
\affiliation{European XFEL, 22869 Schenefeld, Germany}
\author{Xuemei Cheng}
\affiliation{Center for Free-Electron Laser Science CFEL, Deutsches Elektronen-Synchrotron DESY, Hamburg 22607, Germany}
\author{Davide Derelli}
\affiliation{Institute for Nanostructure and Solid-State Physics, Center for Hybrid Nanostructures, University of Hamburg, Hamburg 22761, Germany}
\author{Armando D. Estillore}
\affiliation{Center for Free-Electron Laser Science CFEL, Deutsches Elektronen-Synchrotron DESY, Hamburg 22607, Germany}
\author{Tjark L. R. Groene}
\affiliation{Institute for Nanostructure and Solid-State Physics, Center for Hybrid Nanostructures, University of Hamburg, Hamburg 22761, Germany}
\author{Lukas V. Haas}
\affiliation{Center for Free-Electron Laser Science CFEL, Deutsches Elektronen-Synchrotron DESY, Hamburg 22607, Germany}
\altaffiliation{The Hamburg Center for Ultrafast Imaging, Hamburg 22761, Germany}
\author{Romain Letrun}
\affiliation{European XFEL, 22869 Schenefeld, Germany}
\author{Chan Kim}
\affiliation{European XFEL, 22869 Schenefeld, Germany}
\author{Jayanath C. P. Koliyadu}
\affiliation{European XFEL, 22869 Schenefeld, Germany}
\author{Abhishek Mall}
\affiliation{Max Planck Institute for the Structure and Dynamics of Matter, Hamburg 22761, Germany}
\author{Parichita Mazumder}
\affiliation{Max Planck Institute for the Structure and Dynamics of Matter, Hamburg 22761, Germany}
\altaffiliation{The Hamburg Center for Ultrafast Imaging, Hamburg 22761, Germany}
\author{Diogo V. M. Melo}
\affiliation{European XFEL, 22869 Schenefeld, Germany}
\author{Adam R. Round}
\affiliation{European XFEL, 22869 Schenefeld, Germany}
\author{Amit K. Samanta}
\affiliation{Center for Free-Electron Laser Science CFEL, Deutsches Elektronen-Synchrotron DESY, Hamburg 22607, Germany}
\altaffiliation{The Hamburg Center for Ultrafast Imaging, Hamburg 22761, Germany}
\author{Abhisakh Sarma}
\affiliation{European XFEL, 22869 Schenefeld, Germany}
\author{Zhou Shen}
\affiliation{Max Planck Institute for the Structure and Dynamics of Matter, Hamburg 22761, Germany}
\author{Xiao Sun}
\affiliation{Deutsches Elektronen-Synchrotron DESY, Hamburg 22607, Germany}
\altaffiliation{Institute of Integrated Natural Science, University of Koblenz, Koblenz 56070, Germany}
\author{Patrik Vagovic}
\affiliation{European XFEL, 22869 Schenefeld, Germany}
\author{Tamme Wollweber}
\affiliation{Max Planck Institute for the Structure and Dynamics of Matter, Hamburg 22761, Germany}
\altaffiliation{The Hamburg Center for Ultrafast Imaging, Hamburg 22761, Germany}
\author{Richard Bean}
\affiliation{European XFEL, 22869 Schenefeld, Germany}
\author{Jochen K{\"u}pper}
\affiliation{Center for Free-Electron Laser Science CFEL, Deutsches Elektronen-Synchrotron DESY, Hamburg 22607, Germany}
\author{Henry N. Chapman}
\affiliation{Center for Free-Electron Laser Science CFEL, Deutsches Elektronen-Synchrotron DESY, Hamburg 22607, Germany}
\altaffiliation{The Hamburg Center for Ultrafast Imaging, Hamburg 22761, Germany}
\alsoaffiliation{Department of Physics, Universität Hamburg, 22761 Hamburg, Germany}
\author{Dorota Koziej}
\affiliation{Institute for Nanostructure and Solid-State Physics, Center for Hybrid Nanostructures, University of Hamburg, Hamburg 22761, Germany}
\altaffiliation{The Hamburg Center for Ultrafast Imaging, Hamburg 22761, Germany}
\email{dorota.koziej@uni-hamburg.de}
\author{Kartik Ayyer}
\affiliation{Max Planck Institute for the Structure and Dynamics of Matter, Hamburg 22761, Germany}
\altaffiliation{The Hamburg Center for Ultrafast Imaging, Hamburg 22761, Germany}
\email{kartik.ayyer@mpsd.mpg.de}

\title[running title]{Single-Particle X-ray Scattering Reveals a High Local Supersaturation of Precursors as the Origin of CoO Assembly Formation}

\abbreviations{SPI, CDI, SAXS}
\keywords{SPI, CDI, SAXS, Single Particle X-ray Scattering, CoO nanocrystal assembly}

\begin{document}
\begin{tocentry}  
\centering
\includegraphics[width=1.0\linewidth]{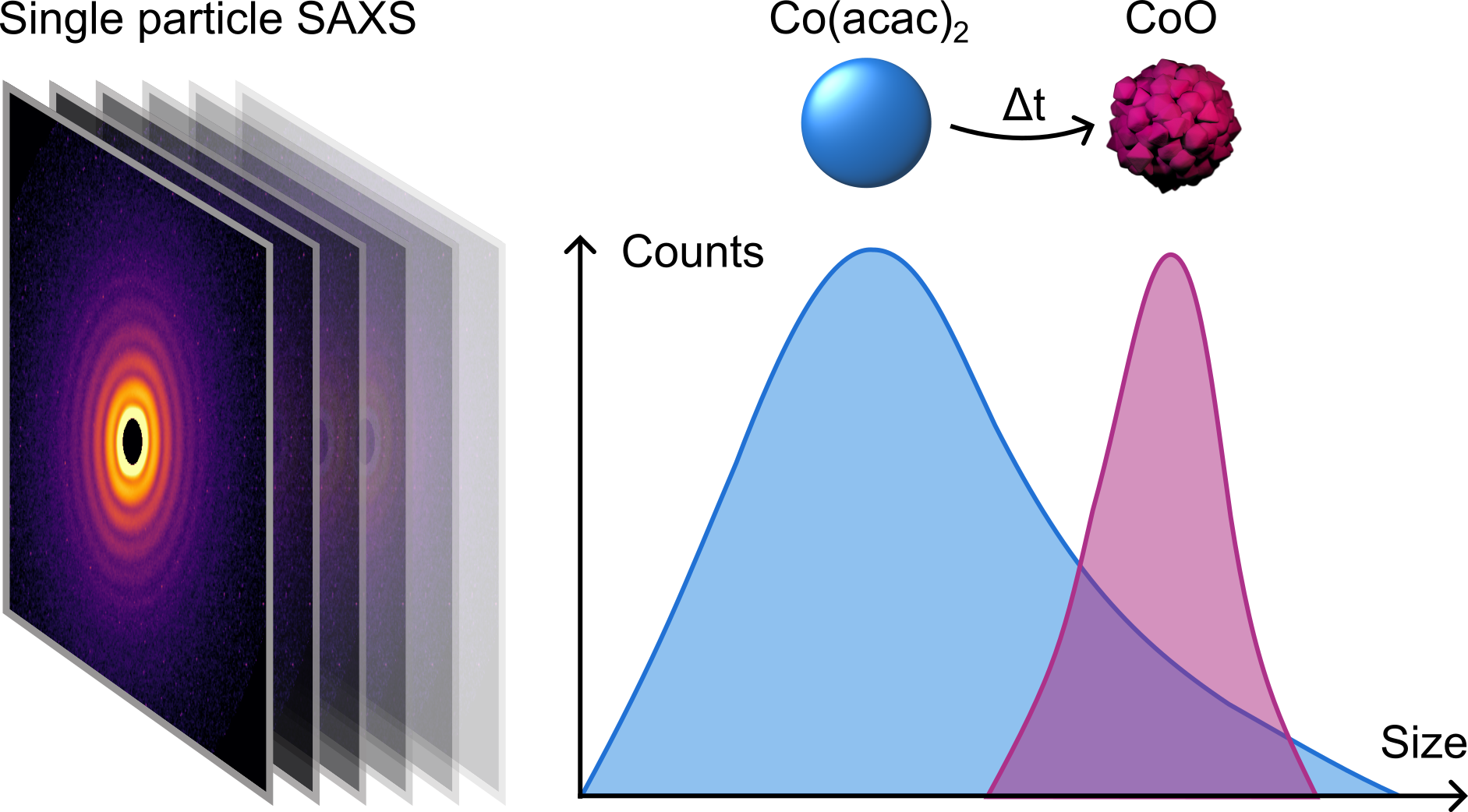}
\end{tocentry}

\begin{abstract}
Single-particle small-angle X-ray scattering (SP-SAXS) enables quantitative morphological analysis by recording diffraction snapshots from isolated particles using X-ray free-electron laser (XFEL) pulses. Unlike conventional X-ray techniques, which average over the entire illuminated sample volume, SP-SAXS resolves low-contrast, less abundant, or transient species within heterogeneous particle populations that would otherwise remain hidden. Here, we apply SP-SAXS to investigate the solvothermal formation of CoO nanocrystal assemblies from a Co(acac)$_3$ precursor in benzyl alcohol. The single-particle data reveal amorphous, uniform-density Co(acac)$_2$ spheres as transient intermediates that directly crystallize into cavernous CoO nanocrystal assemblies, which explains why CoO forms as hierarchical aggregates rather than as isolated nanocrystals. These results demonstrate that SP-SAXS provides a powerful framework for disentangling morphological heterogeneity in nanoparticle formation processes.
\end{abstract}

The emergence of nanomaterials in solution is governed by complex chemical and structural transformations that ultimately dictate their composition, structure, morphology, and functionality. The rational design of nanomaterials with tailored properties therefore requires mechanistic insight into their formation pathways.\cite{heiligtag_fascinating_2013, wu_nucleation_2022, fu_recent_2022} In many systems, nanomaterials do not form through the straightforward monomer-by-monomer growth described by classical nucleation theory but rather follow nonclassical pathways involving metastable intermediates such as pre-nucleation clusters, dense liquid phases, amorphous precipitates, or the assembly of nanoscale building blocks into hierarchical architectures.\cite{lee_nonclassical_2016, gebauer_classical_2018, jun_classical_2022} These multi-step routes have been reported across a wide range of material classes, yet they continue to pose significant challenges for mechanistic understanding and predictive control.\cite{gebauer_crystal_2022, du_non-classical_2024, finney_molecular_2024, jia_new_2024, wang_research_2022}

Among the most powerful methods for investigating nanomaterial formation are X-ray techniques at synchrotron sources.\cite{bojesen_chemistry_2016} For instance, wide-angle X-ray scattering (WAXS) provides access to atomic arrangements, while small-angle X-ray scattering (SAXS) probes particle size, shape, and morphology.\cite{leffler_nanoparticle_2021, whitehead_particle_2021} X-ray absorption spectroscopy (XAS) offers element-specific insight into the electronic structure and chemical environment of the absorbing atom.\cite{koziej_revealing_2016} Complementary optical spectroscopies such as ultraviolet, visible, and infrared (UV/Vis/IR) spectroscopy are sensitive to organic species, optical band-gap transitions, and plasmonic resonances.\cite{bols_situ_2024} Similar to SAXS, dynamic light scattering (DLS) probes the particle size, but it assumes a hard sphere model and is not applicable to broad or multimodal size distributions.\cite{jia_dynamic_2023} Such methods provide comprehensive information about nanoparticle formation and can be applied \textit{in situ}, enabling real-time monitoring of the evolution of the electronic, atomic, and mesoscopic structure. However, they inherently average over the illuminated sample volume, which may obscure structural or chemical heterogeneity within particle ensembles.\cite{li_situ_2022}
The analytical ultracentrifugation (AUC) enables the deconvolution of particle size distributions from sedimentation profiles of colloidal nanoparticle dispersions. However, it relies on assumptions about particle density, shape, and frictional ratio, and thus cannot accurately resolve heterogeneous, complex, or anisotropic morphologies.\cite{colfen_analytical_2023} 

In contrast, individual particles can be directly imaged during formation in solution using \textit{in situ} electron microscopy (EM) or from quenched aliquots via cryogenic (cryo-)EM, which, however, require elaborate sample preparation, and are prone to electron-beam-induced damage and confinement effects, and only very small sample volumes can be probed.\cite{ilett_studying_2024} Similarly, atomic force microscopy (AFM) can resolve surface morphology and size distributions of deposited nanoparticles but is limited to dried samples and small surface areas. In summary, all conventional methods which allow the study of nanomaterial formation mechanisms either lose information by averaging over the whole sample volume, or only allow very small sample quantities and might be further altered due to sample preparation or beam-damage.

Here, we introduce single-particle small-angle X-ray scattering (SP-SAXS), which enables the morphological analysis of very large numbers of individual particles using an X-ray free-electron laser (XFEL). The ultrashort and extremely intense XFEL pulses used in SP-SAXS ensure that diffraction is recorded before the onset of X-ray–induced damage, effectively capturing an undistorted structural snapshot of each particle.\cite{chapman_femtosecond_2011}

We apply SP-SAXS to a model system, the solvothermal synthesis of CoO nanocrystal assemblies from a Co(acac)$_3$ precursor in benzyl alcohol at 160~°C. Previous complementary \textit{in situ} X-ray studies followed the reaction from the molecular precursor to the final assemblies by combining XAS with WAXS and SAXS.\cite{grote_x-ray_2021} XAS revealed the rapid reduction of Co$^{3+}$ to Co$^{2+}$ and identified Co(acac)$_2$ as a stable intermediate, which gradually transformed into rock-salt CoO. Time-resolved WAXS and SAXS analyses showed that crystallite and assembly growth proceeded concurrently, with CoO nanocrystals expanding from $\sim$3~nm to $\sim$6~nm and the corresponding spherical assemblies from $\sim$20~nm to $\sim$60~nm over the course of the reaction. This work provided a comprehensive picture of the chemical reduction, nucleation, and growth steps. However, one central question remained unresolved: why does CoO emerge and grow as an assembly? The data revealed an interconnected evolution of nanocrystals and assemblies but could not disentangle whether the assemblies originate from the crystallization of amorphous intermediates or from particle aggregation, due to the averaging over entire reaction volume of the applied \textit{in situ} X-ray methods.
By analyzing scattering patterns from individual CoO assemblies and pre-assembly entities extracted from the reaction solution during the early stages of assembly formation, we identify uniform amorphous uniform-density spheres as transient intermediates, that subsequently crystallize into cavernous superstructures. This single-particle perspective provides the missing mechanistic link and explains why CoO forms as assemblies rather than as dispersed nanocrystals.

\begin{figure}[H]
    \centering
    \includegraphics[width=1\linewidth]{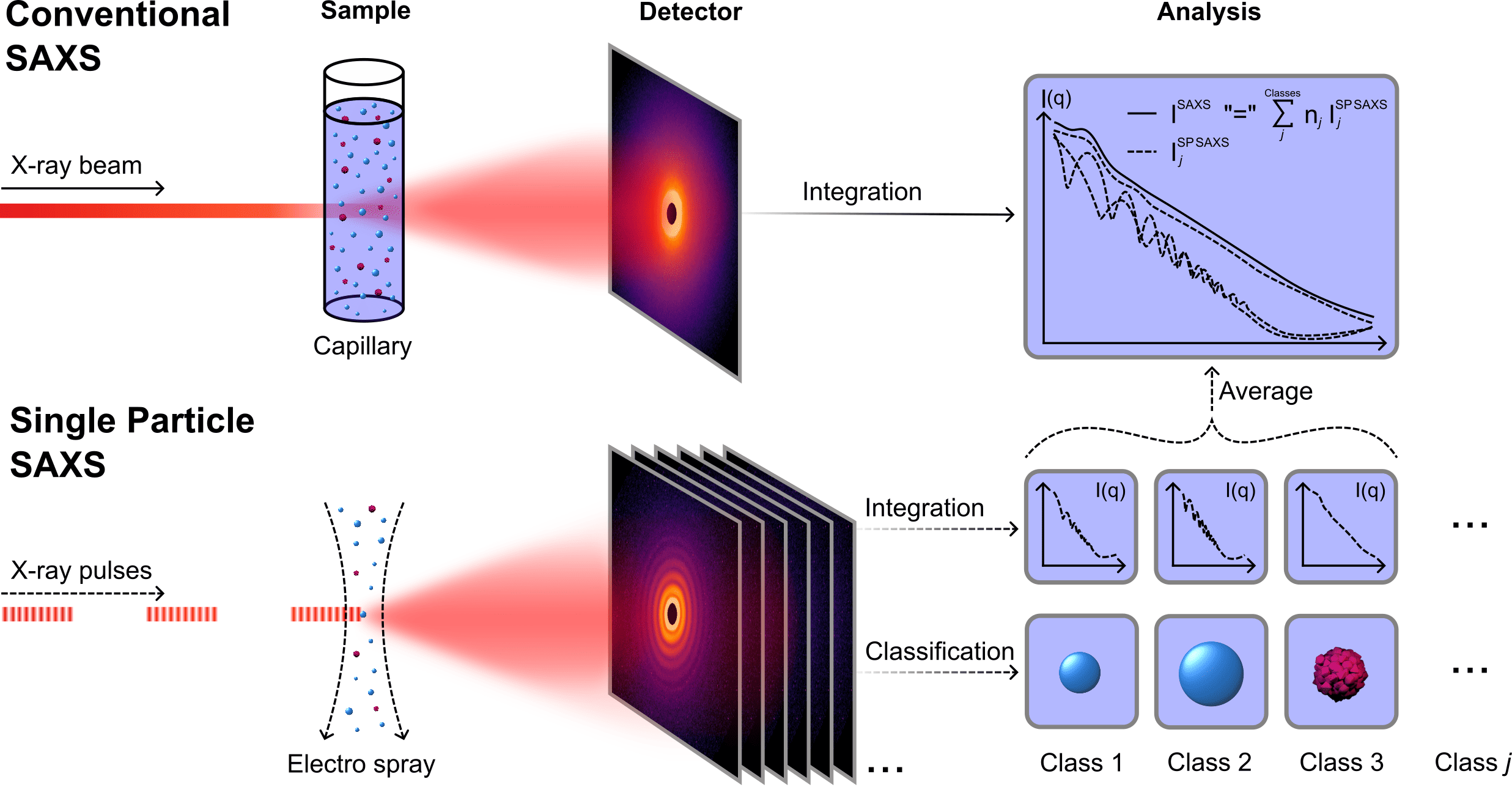}
    \caption{Experimental and analytical workflow of single particle SAXS (SP-SAXS) compared to conventional SAXS. In conventional SAXS, the diffraction pattern comprises scattering contributions of all species within the illuminated sample volume of the X-ray beam from a synchrotron or laboratory source. In SP-SAXS, diffraction patterns from individual particles are averaged into classes, each representing a distinct particle population within the sample. The relative hit ratio of each class, $\mathrm{n}_j$, reflects the concentration of the corresponding particle species $j$. In principle, the sum of all SP-SAXS class diffraction patterns, $\mathrm{I}_j^{\mathrm{SP\text{-}SAXS}}$, reproduces the total diffraction pattern obtained in conventional SAXS, $\mathrm{I}^{\mathrm{SAXS}}$.}
    \label{fig:experimental}
\end{figure}

Figure \ref{fig:experimental} illustrates the experimental and analytical workflow of conventional SAXS in comparison with SP-SAXS. In conventional SAXS, measured at a synchrotron or laboratory X-ray source, each diffraction pattern represents the sum of scattering contributions from all species within the illuminated sample volume. In contrast, SP-SAXS collects diffraction patterns from individual particles that are delivered in a dilute aerosol or liquid jet at an XFEL. The single-particle diffraction patterns are typically noisy, incomplete, and un-oriented. To obtain high-resolution data, a large ensemble of similar single-particle diffraction patterns are identified, orientationally aligned, and averaged. Each averaged dataset forms a class, whose relative hit ratio reflects the population of the corresponding particle type within the sample. The SP-SAXS data processing routine follows similar principles to single-particle imaging (SPI) or coherent diffractive imaging (CDI), in which the individual diffraction patterns are mapped in three-dimensional diffraction space and phase reconstructed.\cite{sobolev_megahertz_2020, ayyer_3d_2021} In SP-SAXS, we analyze the averaged two-dimensional diffraction images and their radial integrations.

To elucidate the CoO nanocrystal assembly formation pathway, we perform SP-SAXS on reaction aliquots collected at three early reaction times during the emergence of the CoO assemblies: 20, 30, and 40 min. In total, we collect \num{650000} single particle diffraction snapshots with an average hit rate of \SI{2.1}{\percent}, from which 60 distinct classes are identified across the combined dataset of the three aliquots. Table \ref{tab:classes} lists all classes including their total hit rate and relative occupancies across the different reaction times. Figures~\ref{fig:si_all_imgs} and \ref{fig:si_all_ri} display the diffraction images and corresponding radial integrations of all classes. 

\begin{figure}[H]
    \centering
    \includegraphics[width=1\linewidth]{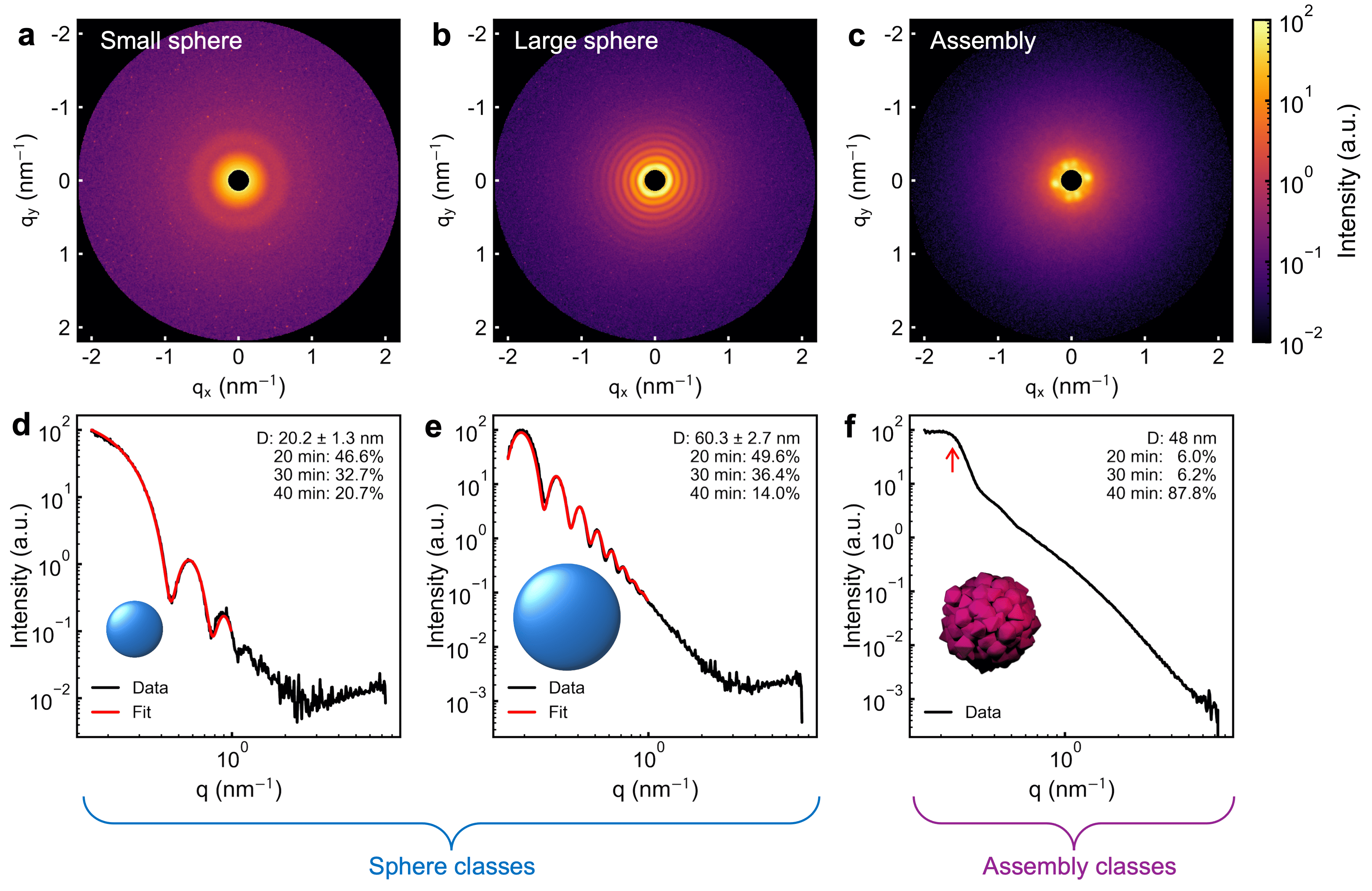}
    \caption{Representative diffraction images (a–c) and corresponding radial integrations (d–f) of selected SP-SAXS classes. The diameter D of the sphere classes is fitted with the a spherical form factor, where the error represents the standard deviation of the Gaussian distribution. The diameter of the assembly classes is estimated by the intensity bump maximum as marked by the red arrow. The relative occupancy of each class at 20, 30, and 40 min reflects the temporal evolution of the populations.}
    \label{fig:example_images_n_ri}
\end{figure}

In Figure~\ref{fig:example_images_n_ri}, we show representative diffraction patterns of selected classes. The scattering profiles can be assigned either to amorphous, uniform-density spheres -- referred to as sphere classes, or to nanocrystal assemblies -- referred to as assembly classes. The sphere classes exhibit isotropic ring patterns in their diffraction images, and the corresponding radial integrations display the characteristic oscillations of monodisperse spherical form factors with an overall $\mathrm{q}^{-4}$ intensity decay. At higher $\mathrm{q}$ values $>$ \SI{3}{\per\nano\meter}, the intensity increases systematically in all sphere classes, which originates from diffuse scattering from the amorphous structure of the spherical particles. The assembly classes, in contrast, display sharp low-$\mathrm{q}$ peaks in the diffraction images, arising from the internal fractal arrangement of nanocrystals within the assemblies. Their radial integrations typically feature one intensity bump around \SI{0.2}{\per\nano\meter}, followed by a smooth decay -- closely resembling the SAXS profile observed after full conversion of the intermediate into CoO assemblies.\cite{grote_x-ray_2021} We estimate the assembly size from the position of the intensity bump. The sphere patterns are modeled using a spherical form factor, incorporating a Gaussian size distribution to account for minor variations in particle size within each sphere class. We note that some sphere classes fit well in the low-$\mathrm{q}$ region of the first fringes, but the model tends to underestimate the intensity for $\mathrm{q} > $~\SI{0.5}{\per\nano\meter}. This weak deviation suggests the onset of structural inhomogeneity, possibly early crystallization within a subset of spheres; however, the effect is subtle and should be regarded as a qualitative trend rather than a quantitative indicator of structural evolution. Synthesis, sample preparation, SP-SAXS data processing, and the fitting procedure and size determination of the classes' diffraction patterns are described in the Supporting Information in detail.

\begin{figure}[H]
    \centering
    \includegraphics[width=1\linewidth]{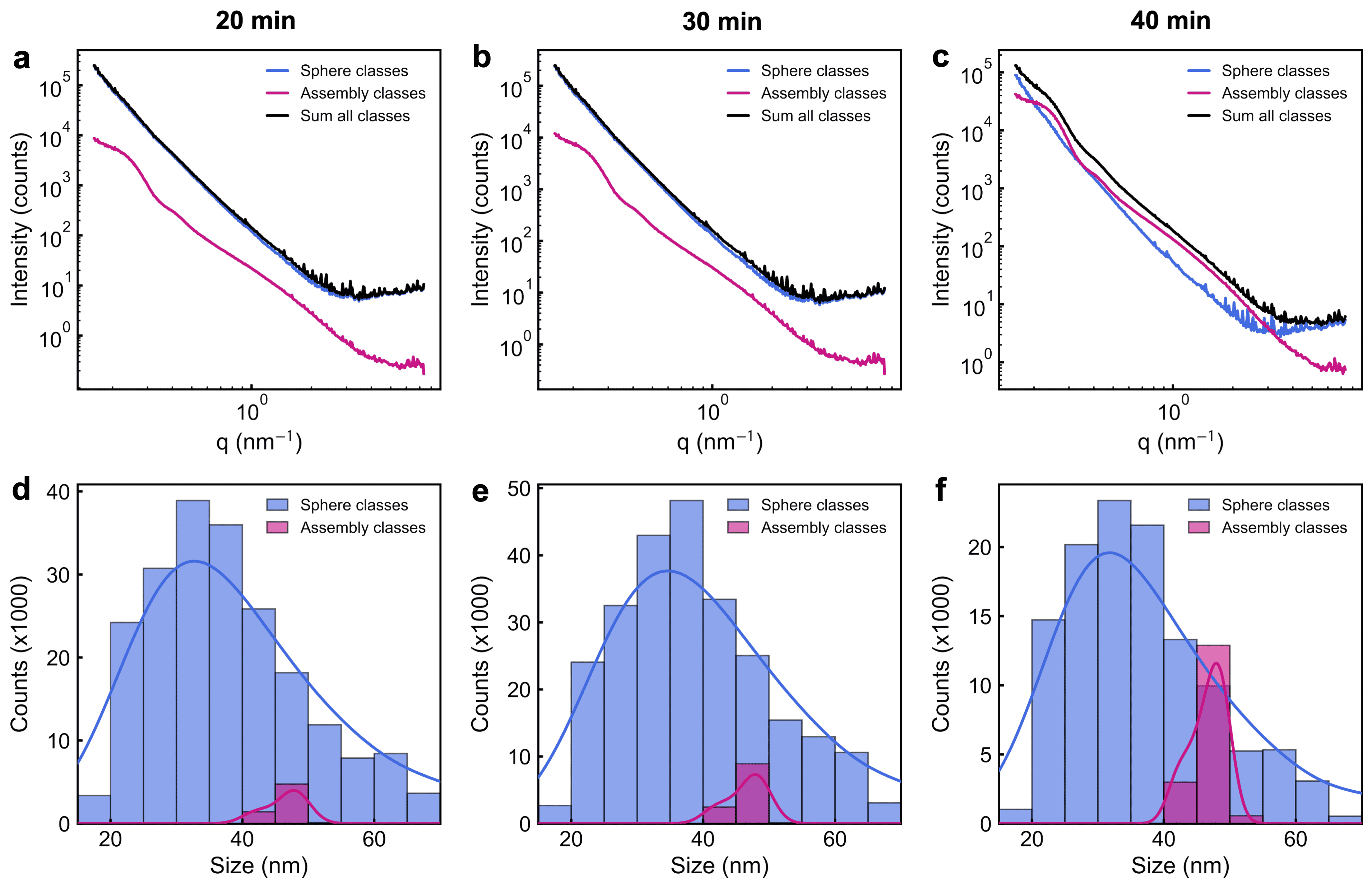}
    \caption{SP-SAXS analysis of the reaction aliquots at 20, 30, and 40 min. (a-c) Summed radial integrations of the sphere, assembly, and all classes. (d-f) Histogram of the size distribution of the sphere and assembly classes. The solid trace shows a kernel density estimate of the histograms.}
    \label{fig:sums_ri_n_size_distribution}
\end{figure}

Figure \ref{fig:sums_ri_n_size_distribution}a-c show the summed radial integrations of the sphere, assembly, and all classes at the different reaction times. The summed sphere classes show a steady $\mathrm{q}^{-4}$ slope, due to smearing of spherical form factor oscillations of the overall polydisperse ensemble, and a positive slope at high $\mathrm{q}$ due to the diffuse scattering of the amorphous spheres. The summed assembly classes radial integrations exhibit the characteristic low $\mathrm{q}$ intensity bump associated with the internal nanocrystal arrangement within the assemblies. At reaction times 20 and 30 min, the scattering contribution from the assembly classes is indistinguishable in the radial integration sum of all classes whereas at 40 min the assembly classes dominate the scattering profile due to its increasing concentration. Figure \ref{fig:sums_ri_n_size_distribution}d–f show the size distributions of the sphere and assembly classes at the respective reaction times, revealing a progressive increase in the fraction of assemblies over time.

\begin{figure}[H]
    \centering
    \includegraphics[width=1\linewidth]{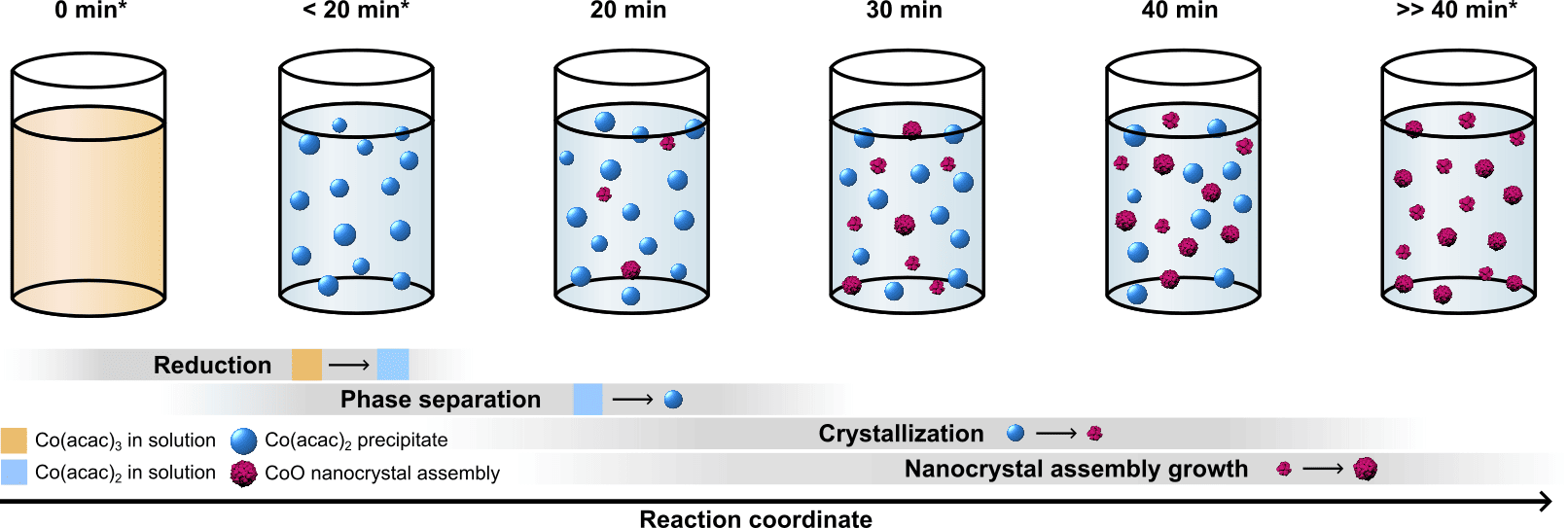}
    \caption{Schematic illustration of the proposed formation pathway of CoO nanocrystal assemblies after the reaction of Co(acac)$_3$ in benzyl alcohol. Initially, Co(acac)$_3$ reduces to Co(acac)$_2$, which subsequently phase-separates to spherical amorphous precipitates. With increasing reaction time, these Co(acac)$_2$ precipitates crystallize into CoO nanocrystal assemblies. The asterisks (*) denote extrapolated reaction states before and after the measured time points of 20, 30, and 40 min.}
    \label{fig:cartoon}
\end{figure}

Altogether, the SP-SAXS analysis reveals a population evolution from amorphous, uniform-density spheres to nanocrystal assemblies, as illustrated in Figure \ref{fig:cartoon}. To interpret these morphological observations, we relate the SP-SAXS results to the chemical transformation pathway established in earlier \textit{in situ} X-ray studies: Initially, the precursor Co(acac)$_3$ is dissolved in benzyl alcohol, where it reduces to the intermediate Co(acac)$_2$, which subsequently transforms into CoO.\cite{grote_x-ray_2021} The SP-SAXS findings indicate that Co(acac)$_2$ phase-separates into spherical amorphous precipitates upon reduction, owing to its low solubility in benzyl alcohol. The comparable size range of these amorphous spheres and the emerging assemblies suggests a direct structural transformation rather than secondary aggregation of individual nanocrystals. Crystallization is likely initiated from high local supersaturation of Co(acac)$_2$ inside the precipitate volume. During crystallization, the higher density of CoO compared to Co(acac)$_2$ causes the spherical precipitates to contract, giving rise to cavernous polycrystalline assemblies instead of dense crystalline entities. 

The Co(acac)$_2$ spheres can easily be overlooked in conventional SAXS measurements, as their smooth intensity decay lacks distinct features in the SAXS regime and may be mistaken for background scattering. In electron microscopy (EM) images, these spheres can also be misinterpreted as organic aggregates or reaction byproducts unrelated to the assembly formation mechanism. Moreover, because of the poor solubility of Co(acac)$_2$ in benzyl alcohol, a significant fraction of the spheres may be lost during sample washing and redispersion. For instance, spherical aggregates are found in the supernatant after washing the reaction mixture with ethanol, as shown in Figure \ref{fig:si_sem_spheres}a,b. This likely explains why the spherical Co(acac)$_2$ particles were not observed in previous studies, where the samples were washed several times with ethanol prior to EM analysis.\cite{grote_x-ray_2021} To further confirm the precipitation behavior of Co(acac)$_2$, Figure \ref{fig:si_sem_spheres}c,d show EM images of commercial Co(acac)$_2$ dissolved in benzyl alcohol and ethanol, both showing precipitation of spherical particles similar to those detected in the reaction solution and supernatant.

\vspace{\baselineskip}
In conclusion, the single-particle perspective provided by SP-SAXS offers unprecedented insight into nanoparticle formation pathways by quantitatively resolving distinct particle populations that may be obscured in conventional measurements. Applied to the solvothermal synthesis of CoO, SP-SAXS reveals that intermediate amorphous Co(acac)$_2$ spheres crystallize into CoO nanocrystal assemblies, elucidating why CoO emerges as hierarchical aggregates rather than as dispersed nanocrystals. Beyond this specific case, SP-SAXS represents a broadly applicable approach for studying morphological and structural heterogeneity in complex systems. Extending the concept to single-particle wide-angle X-ray scattering (SP-WAXS), achieved by reducing the sample-to-detector distance, would allow quantitative access to atomic-scale order similar to serial femtosecond crystallography (SFX).\cite{barends_serial_2022} A multimodal two-detector configuration could further combine SP-SAXS and SP-WAXS, bridging the full range from atomic to mesoscopic structure. Looking ahead, the realization of \textit{in situ} SP-SAXS and SP-WAXS experiments, where small volumes of the reaction mixture are continuously injected into the XFEL beam, will open the way toward real-time visualization of nanoparticle nucleation and growth at the single-particle level, transforming our ability to directly observe matter in formation.

\section*{Author Contributions}
K.A. and D.K. conceived of the project; S.Y.H.-M., L.K., C.A..Z., D.D. and T.L.R.G. prepared the samples under the supervision of D.K.; K.A. coordinated the XFEL experiment and conducted it with A.M., P.M., Z.S., T.W., S.Y.H.-M., L.K., D.D., T.L.R.G., D.K., A.K.S., A.D.E., X.C., L.V.H., J.K., H.N.C., J.B., R.L., C.K., J.C.P.K., D.V.M.M>, A.R.R., A.S., P.V. and R.B.; S.Y.H.-M. and K.A. analyzed the XFEL data with the help of A.M., P.M., Z.S. and T.W.; S.Y.H.-M. collected and analyzed the synchrotron SAXS and WAXS data with the help of X.S.

\begin{acknowledgement}
This research was supported by the Deutsche Forschungsgemeinschaft (DFG) through the Cluster of Excellence “Advanced Imaging of Matter” (EXC 2056, project ID 390715994) and by the European Research Council (LINCHPIN project, grant no. 818941). We acknowledge European XFEL in Schenefeld, Germany, for provision of X-ray free-electron laser beamtime at SPB/SFX SASE1 under proposal number 2995 as well as DESY (Hamburg, Germany), a member of the Helmholtz Association, for the provision of synchrotron beamtime at beamline P21.1\cite{v_zimmermann_p211_2025} and P62. We thank the beamline staffs for the support with the experiments: at P62, Dr. Sylvio Haas; at P21.1, Dr. Martin v. Zimmermann, Dr. Ann-Christin Dippel, Dr. Fernando Igoa, Dr. Jiatu Liu, Philipp Glaevecke, and Olof Gutowski. Further, we thank Stefan Werner from University of Hamburg for the TEM measurements, and we acknowledge financial support from the Open Access Publication Fund of University of Hamburg.
\end{acknowledgement}

\begin{suppinfo}
\renewcommand{\thesection}{S\thechapter}
\renewcommand{\thesubsection}{S\thechapter\arabic{subsection}}
\renewcommand{\thefigure}{S\arabic{figure}}
\renewcommand{\thetable}{S\arabic{table}}

\setcounter{section}{0}
\setcounter{subsection}{0}
\setcounter{figure}{0}
\setcounter{table}{0}

\begin{figure}[H]
    \centering
    \includegraphics[width=0.9\linewidth]{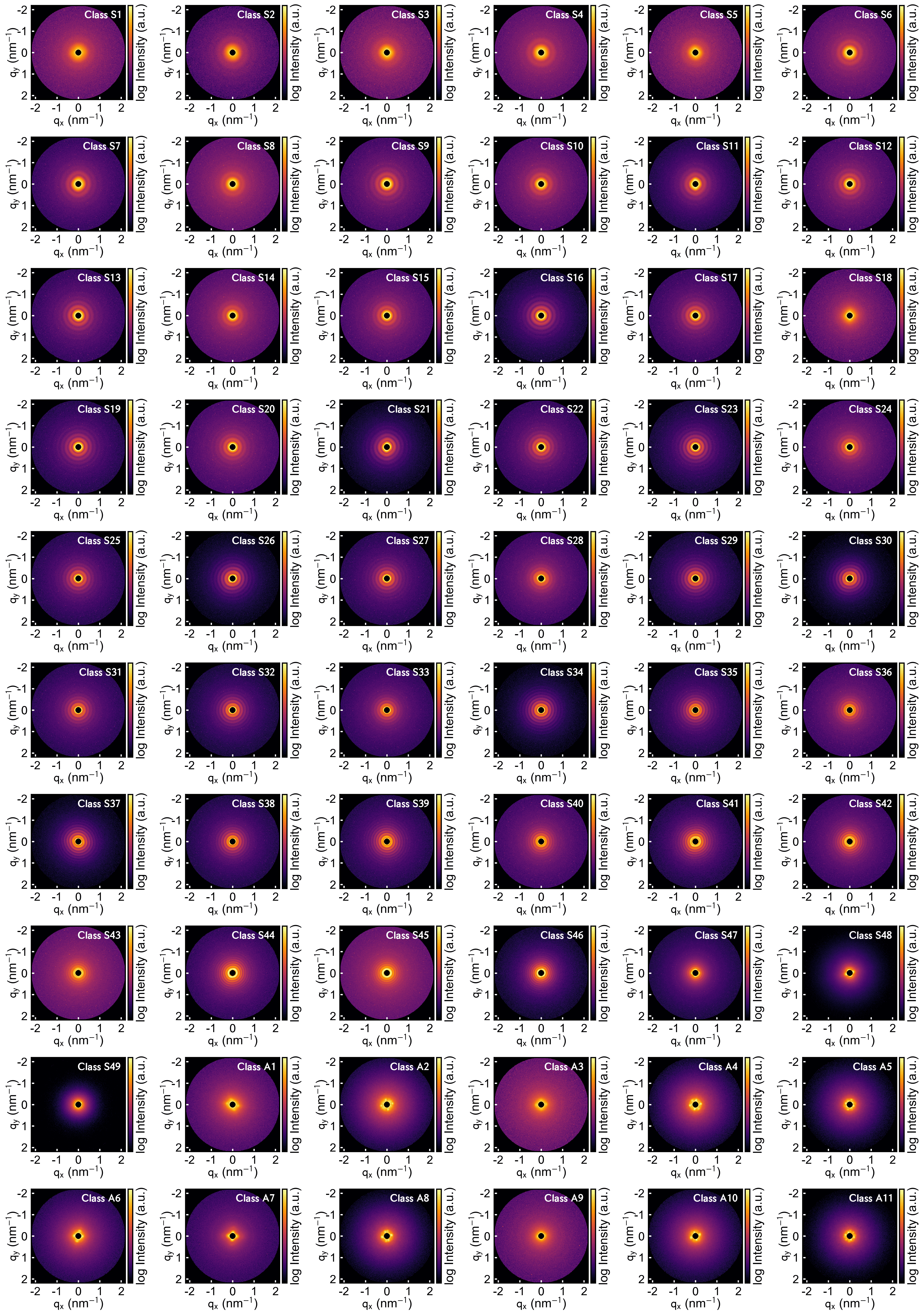}
    \caption{Diffraction images of all SP-SAXS classes.}
    \label{fig:si_all_imgs}
\end{figure}

\begin{figure}[H]
    \centering
    \includegraphics[width=0.9\linewidth]{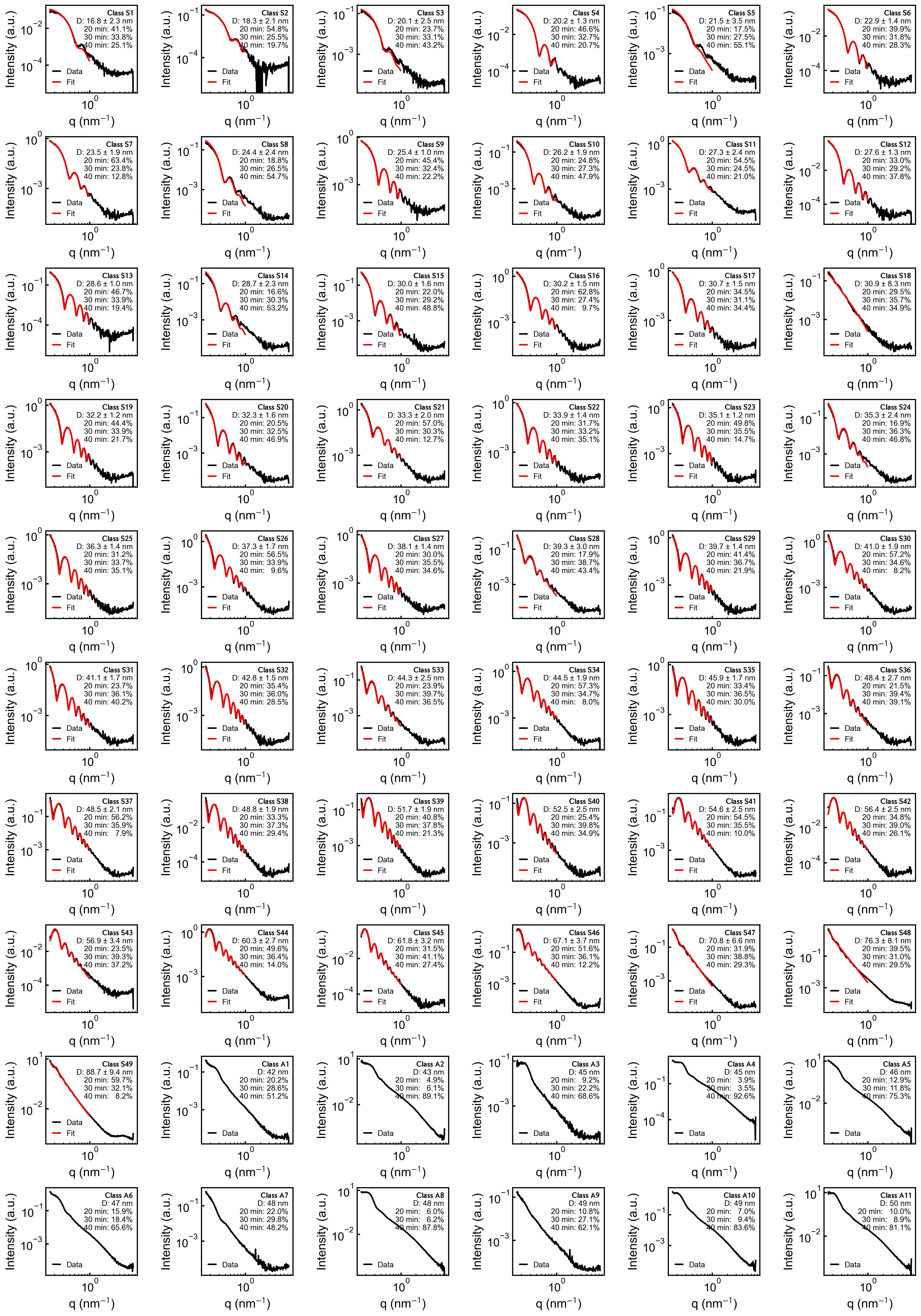}
    \caption{Radial integration of all SP-SAXS classes including their determined diameter D and relative occupancy of the measured reaction times 20, 30, and 40 min.}
    \label{fig:si_all_ri}
\end{figure}

\begin{table}[H]
  \caption{List of all SP-SAXS classes with class number (\#), their classification into sphere (S) or assembly (A) classes, determined particle sizes, relative occupancies at the different reaction times (20, 30, and 40~min), total number of hits per class, and the hit ratio of each class relative to the total number of hits across all classes.}
  \label{tab:classes}
  \scriptsize
  \begin{tabular}{crrrrrr}
    \hline
    \# & Diameter (nm) & 20 min (\%) & 30 min (\%) & 40 min (\%) & No. hits & Hit ratio (\%)\\
    \hline
    S1 & 16.8 $\pm$ 2.3 & 41.1 & 33.8 & 25.1 & 4333 & 0.0067\\
    S2 & 18.3 $\pm$ 2.1 & 54.8 & 25.5 & 19.7 & 2771 & 0.0043\\
    S3 & 20.1 $\pm$ 2.5 & 23.7 & 33.1 & 43.2 & 3280 & 0.0051\\
    S4 & 20.2 $\pm$ 1.3 & 46.6 & 32.7 & 20.7 & 15237 & 0.0237\\
    S5 & 21.5 $\pm$ 3.5 & 17.5 & 27.5 & 55.1 & 3090 & 0.0048\\
    S6 & 22.9 $\pm$ 1.4 & 39.9 & 31.8 & 28.3 & 14541 & 0.0226\\
    S7 & 23.5 $\pm$ 1.9 & 63.4 & 23.8 & 12.8 & 8776 & 0.0137\\
    S8 & 24.4 $\pm$ 2.4 & 18.8 & 26.5 & 54.7 & 18100 & 0.0282\\
    S9 & 25.4 $\pm$ 1.0 & 45.4 & 32.4 & 22.2 & 13540 & 0.0211\\
    S10 & 26.2 $\pm$ 1.9 & 24.8 & 27.3 & 47.9 & 16331 & 0.0254\\
    S11 & 27.3 $\pm$ 2.4 & 54.5 & 24.5 & 21.0 & 9984 & 0.0155\\
    S12 & 27.6 $\pm$ 1.3 & 33.0 & 29.2 & 37.8 & 14975 & 0.0233\\
    S13 & 28.6 $\pm$ 1.0 & 46.7 & 33.9 & 19.4 & 12581 & 0.0196\\
    S14 & 28.7 $\pm$ 2.3 & 16.6 & 30.3 & 53.2 & 16025 & 0.0249\\
    S15 & 30.0 $\pm$ 1.6 & 22.0 & 29.2 & 48.8 & 16310 & 0.0254\\
    S16 & 30.2 $\pm$ 1.5 & 62.8 & 27.4 & 9.7 & 9894 & 0.0154\\
    S17 & 30.7 $\pm$ 1.5 & 34.5 & 31.1 & 34.4 & 16396 & 0.0255\\
    S18 & 30.9 $\pm$ 8.3 & 29.5 & 35.7 & 34.9 & 4160 & 0.0065\\
    S19 & 32.2 $\pm$ 1.2 & 44.4 & 33.9 & 21.7 & 13113 & 0.0204\\
    S20 & 32.3 $\pm$ 1.6 & 20.5 & 32.5 & 46.9 & 17880 & 0.0278\\
    S21 & 33.3 $\pm$ 2.0 & 57.0 & 30.3 & 12.7 & 9569 & 0.0149\\
    S22 & 33.9 $\pm$ 1.4 & 31.7 & 33.2 & 35.1 & 17933 & 0.0279\\
    S23 & 35.1 $\pm$ 1.2 & 49.8 & 35.5 & 14.7 & 12412 & 0.0193\\
    S24 & 35.3 $\pm$ 2.4 & 16.9 & 36.3 & 46.8 & 16678 & 0.0260\\
    S25 & 36.3 $\pm$ 1.4 & 31.2 & 33.7 & 35.1 & 17124 & 0.0267\\
    S26 & 37.3 $\pm$ 1.7 & 56.5 & 33.9 & 9.6 & 12365 & 0.0192\\
    S27 & 38.1 $\pm$ 1.4 & 30.0 & 35.5 & 34.6 & 16589 & 0.0258\\
    S28 & 39.3 $\pm$ 3.0 & 17.9 & 38.7 & 43.4 & 16164 & 0.0252\\
    S29 & 39.7 $\pm$ 1.4 & 41.4 & 36.7 & 21.9 & 14393 & 0.0224\\
    S30 & 41.0 $\pm$ 1.9 & 57.2 & 34.6 & 8.2 & 9600 & 0.0149\\
    S31 & 41.1 $\pm$ 1.7 & 23.7 & 36.1 & 40.2 & 18259 & 0.0284\\
    S32 & 42.8 $\pm$ 1.5 & 35.4 & 36.0 & 28.5 & 15901 & 0.0248\\
    S33 & 44.3 $\pm$ 2.5 & 23.9 & 39.7 & 36.5 & 19276 & 0.0300\\
    S34 & 44.5 $\pm$ 1.9 & 57.3 & 34.7 & 8.0 & 9581 & 0.0149\\
    S35 & 45.9 $\pm$ 1.7 & 33.4 & 36.5 & 30.0 & 15587 & 0.0243\\
    S36 & 48.4 $\pm$ 2.7 & 21.5 & 39.4 & 39.1 & 15322 & 0.0238\\
    S37 & 48.5 $\pm$ 2.1 & 56.2 & 35.9 & 7.9 & 8394 & 0.0131\\
    S38 & 48.8 $\pm$ 1.9 & 33.3 & 37.3 & 29.4 & 13871 & 0.0216\\
    S39 & 51.7 $\pm$ 1.9 & 40.8 & 37.8 & 21.3 & 10418 & 0.0162\\
    S40 & 52.5 $\pm$ 2.5 & 25.4 & 39.8 & 34.9 & 15325 & 0.0239\\
    S41 & 54.6 $\pm$ 2.5 & 54.5 & 35.5 & 10.0 & 6836 & 0.0106\\
    S42 & 56.4 $\pm$ 2.5 & 34.8 & 39.0 & 26.1 & 13104 & 0.0204\\
    S43 & 56.9 $\pm$ 3.4 & 23.5 & 39.3 & 37.2 & 13076 & 0.0204\\
    S44 & 60.3 $\pm$ 2.7 & 49.6 & 36.4 & 14.0 & 8262 & 0.0129\\
    S45 & 61.8 $\pm$ 3.2 & 31.5 & 41.1 & 27.4 & 13832 & 0.0215\\
    S46 & 67.1 $\pm$ 3.7 & 51.6 & 36.1 & 12.2 & 7288 & 0.0113\\
    S47 & 70.8 $\pm$ 6.6 & 31.9 & 38.8 & 29.3 & 17887 & 0.0278\\
    S48 & 76.3 $\pm$ 8.1 & 39.5 & 31.0 & 29.5 & 10107 & 0.0157\\
    S49 & 88.7 $\pm$ 9.4 & 59.7 & 32.1 & 8.2 & 1777 & 0.0028\\
    A1 & 42 & 20.2 & 28.6 & 51.2 & 5813 & 0.0090\\
    A2 & 43 & 4.9 & 6.1 & 89.1 & 1054 & 0.0016\\
    A3 & 45 & 9.2 & 22.2 & 68.6 & 3459 & 0.0054\\
    A4 & 45 & 3.9 & 3.5 & 92.6 & 814 & 0.0013\\
    A5 & 46 & 12.9 & 11.8 & 75.3 & 1103 & 0.0017\\
    A6 & 47 & 15.9 & 18.4 & 65.6 & 3581 & 0.0056\\
    A7 & 48 & 22.0 & 29.8 & 48.2 & 9831 & 0.0153\\
    A8 & 48 & 6.0 & 6.2 & 87.8 & 938 & 0.0015\\
    A9 & 49 & 10.8 & 27.1 & 62.1 & 4950 & 0.0077\\
    A10 & 49 & 7.0 & 9.4 & 83.6 & 1876 & 0.0029\\
    A11 & 50 & 10.0 & 8.9 & 81.1 & 798 & 0.0012\\
    $\sum$ & - & 35.3 & 42.9 & 21.8 & 642461 & 100.0000\\
    \hline
  \end{tabular}
\end{table}

\section*{Experimental}
\textit{Single particle small-angle X-ray scattering (SP-SAXS)}:\newline
The SP-SAXS measurements were performed at the Single Particle, Biomolecules and Clusters/Serial Femtosecond Crystallography (SPB/SFX) end-station at the European X-ray Free Electron Laser (EuXFEL)~\cite{mancuso_single_2019}. X-ray pulses with photon energy of \SI{6}{\kilo\eV} and average pulse energy of \SI{1.2}{\micro\joule} were focused to a diameter of around \SI{250}{\nano\meter}. The sample dispersion was aerosolized and transported to the X-ray interaction region using an electrospray and aerodynamic lens stack injection system~\cite{bielecki_electrospray_2019}. Diffraction patterns were collected at an average rate of 3420 frames/second in 10 bursts of 342 frames per second on the AGIPD-1M detector~\cite{allahgholi_megapixels_2019} placed \SI{700}{\milli\meter} downstream of the interaction point. 

An average of \SI{2.0}{\percent} of the patterns contained statistically significant diffraction from single particles above the background scattering, primarily from the carrier gas. Of these \num{794902} patterns, \num{150200} were discarded due to instabilities in the electrospray, during which very large droplets were produced. The other \num{644702} patterns were first converted to photons using previously described procedures~\cite{ayyer_3d_2021}, and then classified into 50 classes using the \emph{Dragonfly} software~\cite{ayyer_dragonfly_2016}. This classification is performed using the EMC algorithm~\cite{loh_reconstruction_2009}, where intensity models on the detector are determined which maximize the likelihood of generating the observed diffraction patterns using a Poisson noise model. The process was repeated from a random initial guess 5 times, yielding very similar results.

The results of this classification are shown in Figure~\ref{fig:si_all_imgs} and Figure~\ref{fig:si_all_ri}, in the classes labeled S1-S49. A second round of classification was performed with all patterns belonging to class averages which deviated from dense spherical particles, the results of which are marked as A1-A11 in the same figures.
 
\textit{Size determination SP-SAXS sphere classes}:\newline
The sphere classes were modeled using a custom Python script. The spherical form factor $f(q,r)$ is given by

\begin{equation}
f(q,r) \;=\; \frac{\sin(qr) - qr\cos(qr)}{(qr)^3},
\label{eq:f}
\end{equation}

where $q$ is the magnitude of the scattering vector and $r = D/2$ the particle radius, and the intensity of monodisperse spherical particles would be

\begin{equation}
I_{\mathrm{mono}}(q) \;\propto\; f^2(q,r).
\label{eq:i_mono}
\end{equation}

To account for size polydispersity within a class, we assume a Gaussian probability density function $p(r_0,\sigma)$ of the particle radii,

\begin{equation}
p(r, r_0, \sigma) \;=\;
\frac{1}{\sqrt{2\pi}\,\sigma} \exp\left[-\frac{(r - r_0)^2}{2\sigma^2}\right],
\label{eq:gaussian_dist}
\end{equation}

where $r_0$ is the mean radius and $\sigma$ is the standard deviation.
The scattering intensity of a polydisperse ensemble, $I_{\mathrm{poly}}(q)$, is then calculated by summing the form-factor contributions of $N$ discrete radii in a $\pm 3\sigma$-range weighted by their probability:

\begin{equation}
I_{\mathrm{poly}}(q) \;\propto\; f^2_{\mathrm{poly}}(q,r_0,\sigma) \;=\; \sum_{n=1}^{N} p(r_n, r_0,\sigma) \cdot f^2(q,r_n).
\label{eq:f_poly}
\end{equation}

To reduce the influence of the high-intensity low-$q$ region during fitting, we minimize the residual $\chi^2$ using relative intensity weights via

\begin{equation}
\chi^2 \;=\; \sum_{i}
\left[\frac{I_i - I_{\mathrm{model}}(q_i)}{I_i}\right]^2.
\label{eq:relative_wls}
\end{equation}

The parameters $r_0$, $\sigma$, and a scale factor were refined by minimizing $\chi^2$ using the Levenberg–Marquardt algorithm implemented in SciPy’s optimize.least\_squares function.\cite{virtanen_scipy_2020}

\textit{Size determination SP-SAXS assembly classes}:\newline
The particle sizes of the assembly classes were estimated by assuming a uniform density for the spherical assemblies, following the approach of Grote et al.\cite{grote_x-ray_2021} The SP-SAXS profiles do not extend into the Guinier regime, which prevents reliable form-factor fitting. Instead, the assembly sizes were obtained directly from the position of the first intensity oscillation in the scattering profile. Specifically, we determine the size from either the maximum or the subsequent local minimum of the first intensity bump, as illustrated in Figure~\ref{fig:si_ri_assemblies}.

The zero of the first derivative of equation \ref{eq:f} and \ref{eq:i_mono} gives the particle diameter of the first intensity oscillation maximum $q_\mathrm{{1,max}}$ and the subsequent minimum $q_\mathrm{{2,min}}$ via:

\begin{equation}
D_{1,\max} = \frac{11.54}{q_{1,\max}},\qquad
D_{2,\min} = \frac{15.45}{q_{2,\min}}.
\label{eq:q1max2minx}
\end{equation}

\begin{figure}[H]
    \centering
    \includegraphics[width=0.75\linewidth]{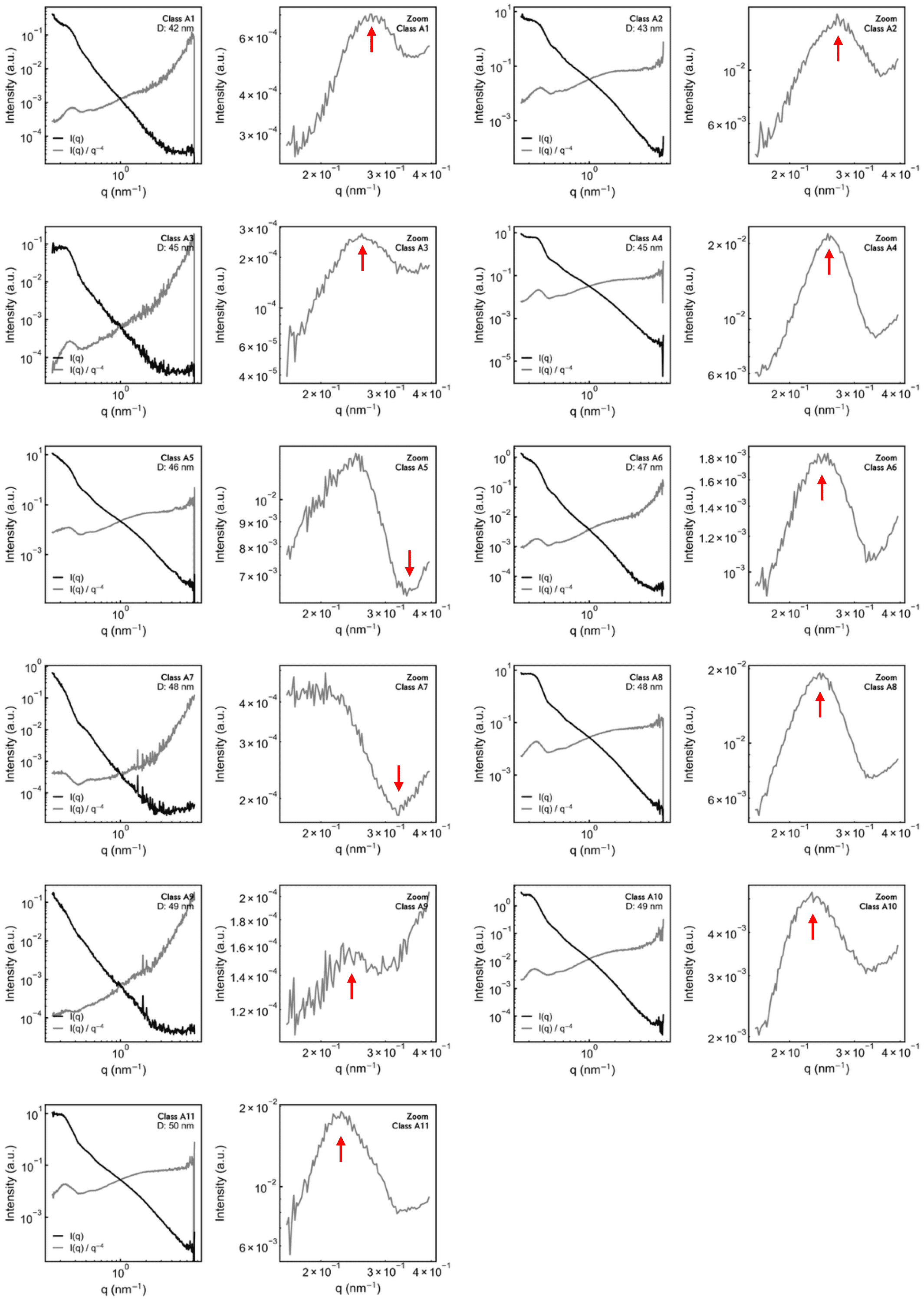}
    \caption{Radial integration of SP-SAXS assembly classes. The upwards facing red arrow indicates the local maximum, $q_{1,\max}$, and the downwards facing red arrow indicated subsequent minimum, $q_{2,\min}$, of the first intensity bump, which was used to determine the size of the assembly. To better see the intensity bump the data is divided by $\mathrm{q^{-4}}$.}
    \label{fig:si_ri_assemblies}
\end{figure}

\textit{Synthesis}:\newline
All chemicals were purchased from commercial sources and used without further purification: Co(acac)$_3$ (Sigma-Aldrich, 99.99\%), Co(acac)$_2$ (Acros, 99.9\%), benzyl alcohol (Sigma-Aldrich, $>$99\%), and ethanol (VWR, absolute grade).\newline
The synthesis is performed as described by Grote et al..\cite{grote_x-ray_2021} Co(acac)$_3$ (179.1~mg, 0.5~mmol) is added to 5~mL of benzyl alcohol and stirred for 10~min at room temperature. 0.8~mL of the reaction solution are then transferred to the reaction container of the reactor, which is described in detail by Grote et.al.. After assembling the reaction container in the reactor, it is first heated to 60~°C with an heating rate of 1~°C/s for 5~min, and then heated to 160~°C with the same heating rate. The time of the beginning of the reaction ($\mathrm{t_0}$) is defined at the point where the heating of the reaction solution from 60~°C to 160~°C starts. All mentions of the reaction times are relative to $\mathrm{t_0}$. For the SP-SAXS measurements the reaction was stopped after 20, 30, and 40~min, and the reactor was cooled with a cold metal block. Figures \ref{fig:si_saxs}, \ref{fig:si_pdf}, and \ref{fig:si_em_assemblies} show conventional SAXS, PDF, and TEM data, respectively, of the 20, 30, and 40~min samples.\newline
The samples for the reference EM measurements of Co(acac)$_2$ precipitates shown in Figure \ref{fig:si_sem_spheres}c,d were prepared by stirring Co(acac)$_2$ (55.4~mg, 0.2~mmol) in 2~mL benzyl alcohol or ethanol for 30~min at room temperature.\newline

\textit{Sample preparation}:\newline
The SP-SAXS samples were prepared by centrifuging the quenched reaction solution in ethanol for 5~min at 3500~rpm, discarding the supernatant, and collecting the residue. The residue was then weighed and redispersed in a 10~mmol ammonium acetate ethanol solution to obtain a concentration of 0.075~mg/mL.\newline
Conventional SAXS samples were prepared either by directly filling the reaction solution in a capillary or by centrifuging at 3500~rpm for 5~min in ethanol, redispersing the residue in ethanol and filling the capillary with the dispersion.\newline
PDF samples were measured from dry powder after centrifuging at 3500~rpm for 5~min in ethanol.\newline
EM samples were prepared by depositing one drop of the sample solution on a TEM grid and washing the grid with a few drops of ethanol.\newline

\begin{figure}[H]
    \centering
    \includegraphics[width=1\linewidth]{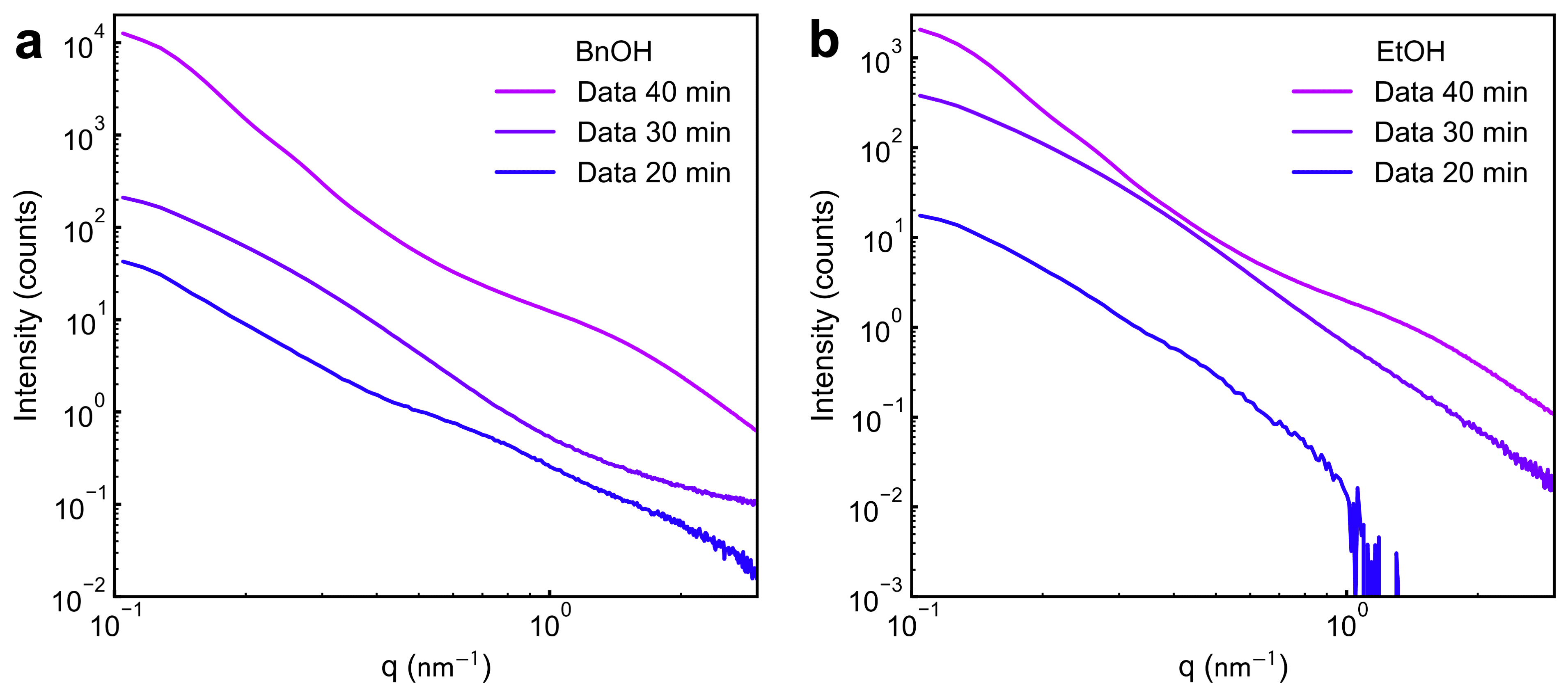}
    \caption{Conventional SAXS data of (a) reaction solutions in benzyl alcohol (BnOH) after 20, 30, and 40~min reaction time and (b) the reaction solutions redispersed in ethanol (EtOH) after centrifuging at 3500~rpm for 5~min. Both the BnOH and EtOH data show similar scattering profiles at the respective time points. Both data sets were measured at the lower part of capillary, which showed sample precipitation, making a comparison of the conventional SAXS to the summed SP-SAXS data, which is measured from very diluted particles, difficult.}
    \label{fig:si_saxs}
\end{figure}

\begin{figure}[H]
    \centering
    \includegraphics[width=0.9\linewidth]{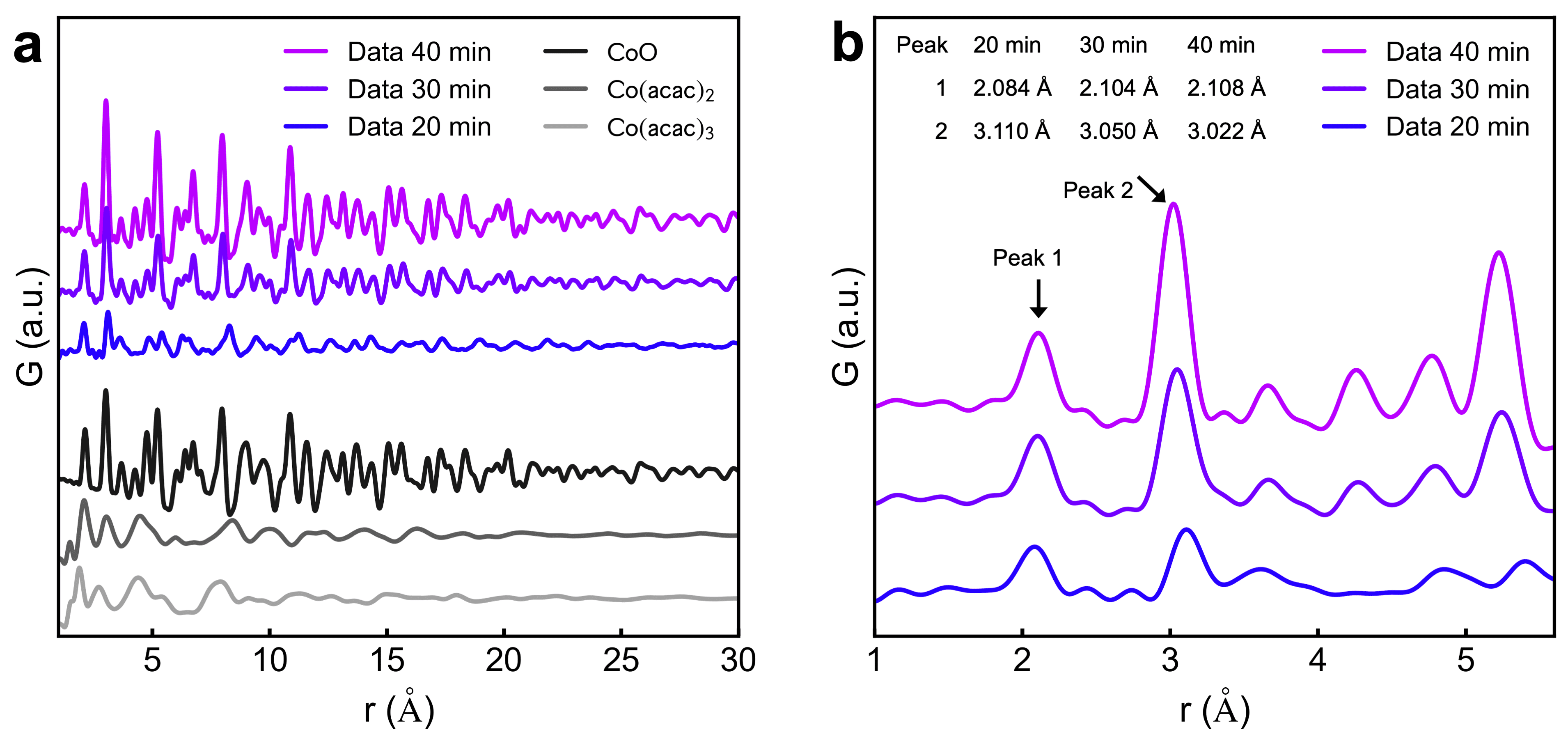}
    \caption{(a) Pair distribution function (PDF) of total X-ray scattering data of 20, 30, and 40~min samples compared to PDF simulations of rock-salt CoO, Co(acac)$_2$\cite{vreshch_monomeric_2010}, and Co(acac)$_3$\cite{chrzanowski_-tris24-pentanedionato-2oocobaltiii_2007} structures. The 30 and 40~min PDF closely matches the CoO simulation, while the 20~min sample shows features of both CoO and Co(acac)$_2$, which confirms the proposed phase transition of Co(acac)$_2$ to CoO. (b) Zoom of the experimental PDFs. The inset lists the peak position of the first and second peak. A detailed PDF analysis is beyond the scope of this paper, but available in the literature.\cite{grote_x-ray_2021}}
    \label{fig:si_pdf}
\end{figure}

\begin{figure}[H]
    \centering
    \includegraphics[width=1.0\linewidth]{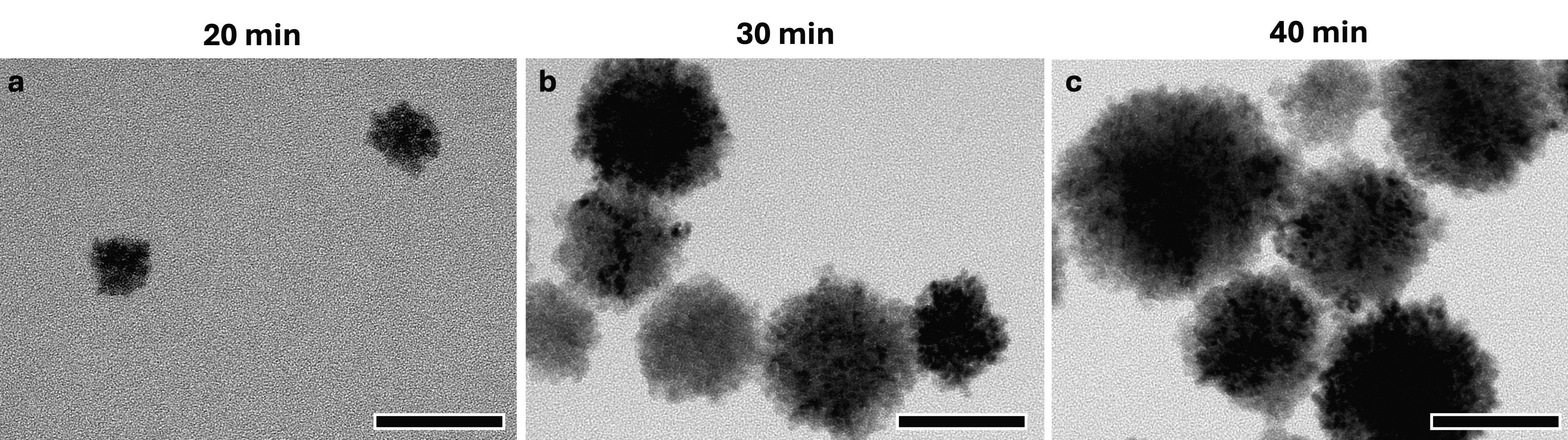}
    \caption{TEM images of CoO nanocrystal assemblies of reaction aliquots of (a) 20~min, (b) 30~min, and (c) 40~min. Scale bars 50~nm.}
    \label{fig:si_em_assemblies}
\end{figure}

\begin{figure}[H]
    \centering
    \includegraphics[width=1.0\linewidth]{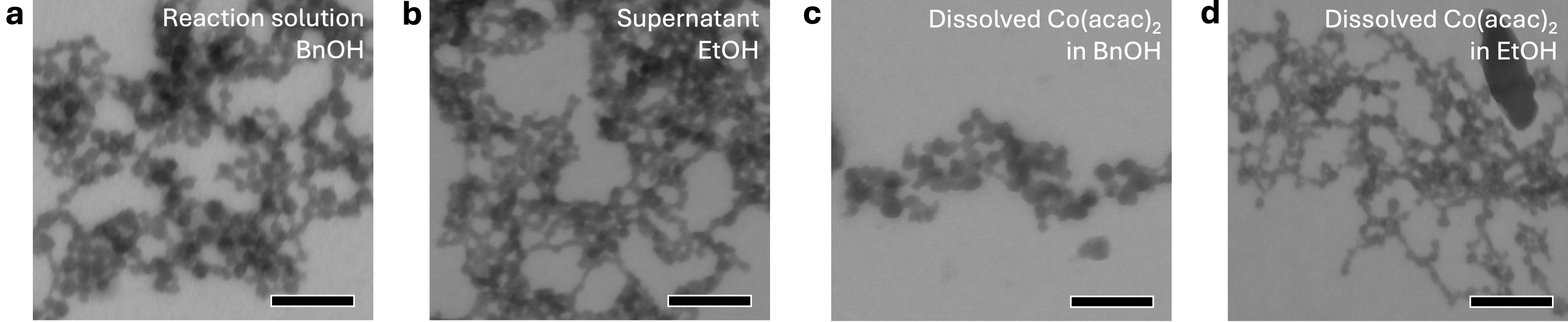}
    \caption{STEM images of spherical aggregates from solutions in benzyl alcohol (BnOH) or ethanol (EtOH). (a) Quenched reaction solution of the reaction of Co(acac)$_3$ in BnOH after 20 min reaction time. (b) Supernatant of washed reaction solution of (a) after centrifugation for 5 min at 3500 rpm in EtOH. Control experiments: commercial Co(acac)$_2$ dissolved in (c) BnOH and (d) EtOH for 30 min. Scale bars: 100~nm.}
    \label{fig:si_sem_spheres}
\end{figure}

\textit{Reproducibility of the synthesis}:\newline
The reaction kinetics observed in this study appear slightly slower than those reported by Grote et al.\cite{grote_x-ray_2021} based on the comparison of SAXS, EM, and PDF data. This difference is related to slight variations in the reactor inlet design, as discussed elsewhere.\cite{harouna-mayer_modular_2025} The overall trends are similar.\newline
Furthermore, we occasionally observe crumpled sheet- or rose-like particles with sizes of approximately 0.5–1.0~$\mathrm{\mu}$m, as shown in Figure \ref{fig:si_em_rose}. These particles are beyond the detection limit of our SP-SAXS set-up and appear more frequently at 20~min, while they are rarely seen at 30 or 40~min. Similar rose-like structures were also reported by Grote et al.\cite{grote_x-ray_2021} at reaction times earlier than 20~min. Given their disappearance as the reaction progresses, we attribute these particles to side reactions that do not affect the formation of CoO nanocrystal assemblies.


\begin{figure}[H]
    \centering
    \includegraphics[width=0.25\linewidth]{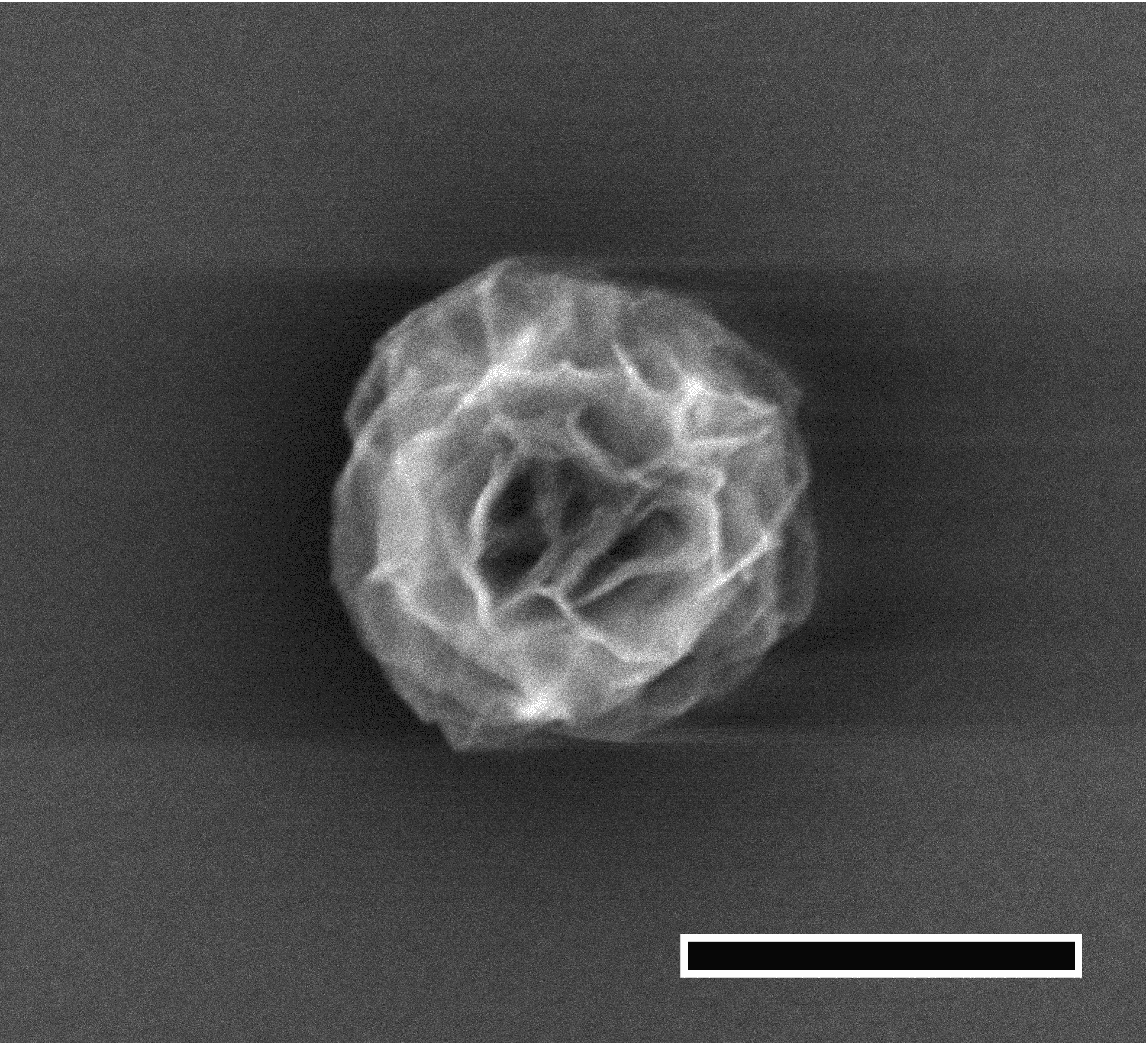}
    \caption{SEM image of a rose-like particle from a 20~min reaction aliquot. Scale bar: 500~nm.}
    \label{fig:si_em_rose}
\end{figure}

\textit{Electron microscopy (EM)}:\newline
Scanning electron microscopy (SEM) and scanning transmission electron microscopy (STEM) images were collected using a Regulus 8220 (Hitachi High Technologies Corp.) with an acceleration voltage of 30~keV. Transmission electron microscopy (TEM) images were collected using a  JEM 1011 (JEOL Ltd.) with an acceleration voltage of 100~keV.

\textit{Conventional SAXS}:\newline
SAXS data was acquired at beamline P62 of PETRA III at Deutsches Elektronen–Synchrotron DESY, Hamburg, Germany. The particle dispersions were filled in a \SI{1}{\milli\meter} diameter borosilicate capillary and diffraction images were recorded for \SI{30}{\second} at an X-ray energy of \SI{24.30}{\kilo\eV} ($\mathrm{\lambda}$ = \SI{0.5102}{\angstrom}) using a two-dimensional X-ray detector (EIGER2 X 9M, Dectris Ltd.) with $3108 \times 3262$ pixels and a pixel size of $75\times$\SI{75}{\micro\meter\squared} and a sample-to-detector distance of \SI{4.946}{\meter}, obtained from a calibration with a silver behenate standard packed into a capillary. The diffraction images were integrated using PyFAI\cite{kieffer_pyfai_2013}.

\textit{Total X-ray scattering (TS) and pair distribution function (PDF) analysis}:\newline
TS data was acquired at beamline P21.1\cite{v_zimmermann_p211_2025} of PETRA III at Deutsches Elektronen–Synchrotron DESY, Hamburg, Germany. The powder sample was packed in a \SI{1}{\milli\meter} diameter borosilicate capillary and diffraction images were recorded for \SI{60}{\second} at an X-ray energy of \SI{101.39}{\kilo\eV} ($\mathrm{\lambda}$ = \SI{0.1222}{\angstrom}) using a two-dimensional X-ray detector (PerkinElmer XRD1621, Varex Imaging Corp.) with $2048\times2048$ pixels and a pixel size of $200\times$\SI{200}{\micro\meter\squared} and a sample-to-detector distance of \SI{0.301}{\meter}, obtained from a calibration with a \ce{LaB6} powder standard packed into a capillary. The diffraction images were integrated using PyFAI\cite{kieffer_pyfai_2013}. The experimental PDFs were calculated using PDFgetX3\cite{juhas_pdfgetx3_2013} with values $\mathrm{q_{max,inst}}=$\SI{25.0}{\angstrom}, $\mathrm{q_{max}}=$\SI{22.5}{\per\angstrom}, $\mathrm{q_{min}}=$\SI{1.0}{\per\angstrom}, and $\mathrm{r_{poly} = 0.9}$. The PDF simulations were calculated using DiffPy-CMI\cite{juhas_complex_2015} with the same values as the experimental PDF calculations.

\end{suppinfo}

\bibliography{references}

@article{grote_x-ray_2021,
	title = {X-ray studies bridge the molecular and macro length scales during the emergence of {CoO} assemblies},
	volume = {12},
	issn = {2041-1723},
	url = {http://www.nature.com/articles/s41467-021-24557-z},
	doi = {10.1038/s41467-021-24557-z},
	pages = {4429},
	number = {1},
	journal = {Nature Communications},
	shortjournal = {Nat Commun},
	author = {Grote, Lukas and Zito, Cecilia A. and Frank, Kilian and Dippel, Ann-Christin and Reisbeck, Patrick and Pitala, Krzysztof and Kvashnina, Kristina O. and Bauters, Stephen and Detlefs, Blanka and Ivashko, Oleh and Pandit, Pallavi and Rebber, Matthias and Harouna-Mayer, Sani Y. and Nickel, Bert and Koziej, Dorota},
	urldate = {2021-07-23},
	date = {2021-12},
    year = {2021},
	langid = {english},
}

@article{heiligtag_fascinating_2013,
	title = {The fascinating world of nanoparticle research},
	volume = {16},
	rights = {https://www.elsevier.com/tdm/userlicense/1.0/},
	issn = {13697021},
	url = {https://linkinghub.elsevier.com/retrieve/pii/S1369702113002253},
	doi = {10.1016/j.mattod.2013.07.004},
	pages = {262--271},
	number = {7},
	journal = {Materials Today},
	shortjournal = {Materials Today},
	author = {Heiligtag, Florian J. and Niederberger, Markus},
	urldate = {2025-06-10},
	date = {2013-07},
    year = {2013},
	langid = {english},
}

@article{jun_classical_2022,
	title = {Classical and Nonclassical Nucleation and Growth Mechanisms for Nanoparticle Formation},
	volume = {73},
	issn = {0066-426X, 1545-1593},
	url = {https://www.annualreviews.org/content/journals/10.1146/annurev-physchem-082720-100947},
	doi = {10.1146/annurev-physchem-082720-100947},
	pages = {453--477},
	issue = {Volume 73, 2022},
	journal = {Annual Review of Physical Chemistry},
	author = {Jun, Young-Shin and Zhu, Yaguang and Wang, Ying and Ghim, Deoukchen and Wu, Xuanhao and Kim, Doyoon and Jung, Haesung},
	urldate = {2025-07-22},
	date = {2022-04-20},
    year = {2022},
	langid = {english},
}

@article{leffler_nanoparticle_2021,
	title = {Nanoparticle Heat-Up Synthesis: In Situ X-ray Diffraction and Extension from Classical to Nonclassical Nucleation and Growth Theory},
	volume = {15},
	issn = {1936-0851},
	url = {https://doi.org/10.1021/acsnano.0c07359},
	doi = {10.1021/acsnano.0c07359},
	shorttitle = {Nanoparticle Heat-Up Synthesis},
	pages = {840--856},
	number = {1},
	journal = {{ACS} Nano},
	shortjournal = {{ACS} Nano},
	author = {Leffler, Vanessa and Ehlert, Sascha and Förster, Beate and Dulle, Martin and Förster, Stephan},
	urldate = {2025-07-22},
	date = {2021-01-26},
    year = {2021},
}

@article{lee_nonclassical_2016,
	title = {Nonclassical nucleation and growth of inorganic nanoparticles},
	volume = {1},
	rights = {https://www.springernature.com/gp/researchers/text-and-data-mining},
	issn = {2058-8437},
	url = {https://www.nature.com/articles/natrevmats201634},
	doi = {10.1038/natrevmats.2016.34},
	number = {8},
	journal = {Nature Reviews Materials},
	shortjournal = {Nat Rev Mater},
	author = {Lee, Jisoo and Yang, Jiwoong and Kwon, Soon Gu and Hyeon, Taeghwan},
	urldate = {2025-07-22},
	date = {2016-06-01},
    year = {2016},
	langid = {english},
}

@article{gebauer_classical_2018,
	title = {On classical and non-classical views on nucleation},
	volume = {318},
	issn = {0002-9599, 1945-452X},
	url = {https://ajsonline.org/article/65749},
	doi = {10.2475/09.2018.05},
	pages = {969--988},
	number = {9},
	journal = {American Journal of Science},
	shortjournal = {Am J Sci},
	author = {Gebauer, Denis and Raiteri, Paolo and Gale, Julian D. and Cölfen, Helmut},
	urldate = {2025-07-22},
	date = {2018-11},
    year = {2018},
	langid = {english},
}

@article{bojesen_chemistry_2016,
	title = {The chemistry of nucleation},
	volume = {18},
	issn = {1466-8033},
	url = {https://xlink.rsc.org/?DOI=C6CE01489E},
	doi = {10.1039/c6ce01489e},
	pages = {8332--8353},
	number = {43},
	journal = {{CrystEngComm}},
	author = {Bøjesen, E. D. and Iversen, B. B.},
	urldate = {2025-07-29},
	date = {2016},
    year = {2016},
	langid = {english},
}

@article{koziej_revealing_2016,
	title = {Revealing Complexity of Nanoparticle Synthesis in Solution by in Situ Hard X-ray Spectroscopy—Today and Beyond},
	volume = {28},
	issn = {0897-4756},
	url = {https://doi.org/10.1021/acs.chemmater.6b00486},
	doi = {10.1021/acs.chemmater.6b00486},
	pages = {2478--2490},
	number = {8},
	journal = {Chemistry of Materials},
	shortjournal = {Chem. Mater.},
	author = {Koziej, Dorota},
	urldate = {2025-06-10},
	date = {2016-04-26},
    year = {2016},
}

@article{ayyer_3d_2021,
	title = {3D diffractive imaging of nanoparticle ensembles using an x-ray laser},
	volume = {8},
	issn = {2334-2536},
	url = {https://www.osapublishing.org/abstract.cfm?URI=optica-8-1-15},
	doi = {10.1364/OPTICA.410851},
	pages = {15},
	number = {1},
	journal = {Optica},
	shortjournal = {Optica},
	author = {Ayyer, Kartik and Xavier, P. Lourdu and Bielecki, Johan and Shen, Zhou and Daurer, Benedikt J. and Samanta, Amit K. and Awel, Salah and Bean, Richard and Barty, Anton and Bergemann, Martin and Ekeberg, Tomas and Estillore, Armando D. and Fangohr, Hans and Giewekemeyer, Klaus and Hunter, Mark S. and Karnevskiy, Mikhail and Kirian, Richard A. and Kirkwood, Henry and Kim, Yoonhee and Koliyadu, Jayanath and Lange, Holger and Letrun, Romain and Lübke, Jannik and Michelat, Thomas and Morgan, Andrew J. and Roth, Nils and Sato, Tokushi and Sikorski, Marcin and Schulz, Florian and Spence, John C. H. and Vagovic, Patrik and Wollweber, Tamme and Worbs, Lena and Yefanov, Oleksandr and Zhuang, Yulong and Maia, Filipe R. N. C. and Horke, Daniel A. and Küpper, Jochen and Loh, N. Duane and Mancuso, Adrian P. and Chapman, Henry N.},
	urldate = {2021-01-28},
	date = {2021-01-20},
    year = {2021},
	langid = {english},
}

@article{sobolev_megahertz_2020,
	title = {Megahertz single-particle imaging at the European {XFEL}},
	volume = {3},
	issn = {2399-3650},
	url = {https://www.nature.com/articles/s42005-020-0362-y},
	doi = {10.1038/s42005-020-0362-y},
	pages = {97},
	number = {1},
	journal = {Communications Physics},
	shortjournal = {Commun Phys},
	author = {Sobolev, Egor and Zolotarev, Sergei and Giewekemeyer, Klaus and Bielecki, Johan and Okamoto, Kenta and Reddy, Hemanth K. N. and Andreasson, Jakob and Ayyer, Kartik and Barak, Imrich and Bari, Sadia and Barty, Anton and Bean, Richard and Bobkov, Sergey and Chapman, Henry N. and Chojnowski, Grzegorz and Daurer, Benedikt J. and Dörner, Katerina and Ekeberg, Tomas and Flückiger, Leonie and Galzitskaya, Oxana and Gelisio, Luca and Hauf, Steffen and Hogue, Brenda G. and Horke, Daniel A. and Hosseinizadeh, Ahmad and Ilyin, Vyacheslav and Jung, Chulho and Kim, Chan and Kim, Yoonhee and Kirian, Richard A. and Kirkwood, Henry and Kulyk, Olena and Küpper, Jochen and Letrun, Romain and Loh, N. Duane and Lorenzen, Kristina and Messerschmidt, Marc and Mühlig, Kerstin and Ourmazd, Abbas and Raab, Natascha and Rode, Andrei V. and Rose, Max and Round, Adam and Sato, Takushi and Schubert, Robin and Schwander, Peter and Sellberg, Jonas A. and Sikorski, Marcin and Silenzi, Alessandro and Song, Changyong and Spence, John C. H. and Stern, Stephan and Sztuk-Dambietz, Jolanta and Teslyuk, Anthon and Timneanu, Nicusor and Trebbin, Martin and Uetrecht, Charlotte and Weinhausen, Britta and Williams, Garth J. and Xavier, P. Lourdu and Xu, Chen and Vartanyants, Ivan A. and Lamzin, Victor S. and Mancuso, Adrian and Maia, Filipe R. N. C.},
	urldate = {2025-10-23},
	date = {2020-05-29},
    year = {2020},
	langid = {english},
}

@article{barends_serial_2022,
	title = {Serial femtosecond crystallography},
	volume = {2},
	issn = {2662-8449},
	url = {https://www.nature.com/articles/s43586-022-00141-7},
	doi = {10.1038/s43586-022-00141-7},
	pages = {59},
	number = {1},
	journal = {Nature Reviews Methods Primers},
	shortjournal = {Nat Rev Methods Primers},
	author = {Barends, Thomas R. M. and Stauch, Benjamin and Cherezov, Vadim and Schlichting, Ilme},
	urldate = {2025-11-01},
	date = {2022-08-04},
    year = {2022},
	langid = {english},
}

@article{harouna-mayer_modular_2025,
	title = {Modular Reactor for In Situ X-ray Scattering, Spectroscopy, and ATR-IR Studies of Solvothermal Nanoparticle Synthesis},
	volume = {n.a.},
	url = {https://arxiv.org/abs/2510.06770},
	doi = {n.a.},
	journal = {arXiv},
	shortjournal = {arXiv},
	author = {Sani Y. Harouna-Mayer and Melike Gumus Akcaalan and Jagadesh Kopula Kesavan and Tjark R. L. Groene and Lars Klemeyer and Sarah-Alexandra Hussak and Lukas Grote and a Davide Derelli and Francesco Caddeo and Cecilia Zito and Paul Stützle and Dorota Speer and Ann-Christin Dippel and Blanka Detlefs and Yannik Appiarius and Axel Jacobi von Wangelin and Dorota Koziej},
    year={2025},
	date = {2025-10-08},
	langid = {english},
}

@article{li_situ_2022,
	title = {\textit{In Situ} Kinetic Observations on Crystal Nucleation and Growth},
	volume = {122},
	rights = {https://doi.org/10.15223/policy-029},
	issn = {0009-2665, 1520-6890},
	url = {https://pubs.acs.org/doi/10.1021/acs.chemrev.1c01067},
	doi = {10.1021/acs.chemrev.1c01067},
	pages = {16911--16982},
	number = {23},
	journal = {Chemical Reviews},
	shortjournal = {Chem. Rev.},
	author = {Li, Junjie and Deepak, Francis Leonard},
	urldate = {2025-11-05},
	date = {2022-12-14},
    year = {2022},
	langid = {english},
}

@article{gebauer_crystal_2022,
	title = {Crystal Nucleation and Growth of Inorganic Ionic Materials from Aqueous Solution: Selected Recent Developments, and Implications},
	volume = {18},
	issn = {1613-6829},
	url = {https://onlinelibrary.wiley.com/doi/abs/10.1002/smll.202107735},
	doi = {10.1002/smll.202107735},
	shorttitle = {Crystal Nucleation and Growth of Inorganic Ionic Materials from Aqueous Solution},
	pages = {2107735},
	number = {28},
	journal = {Small},
	author = {Gebauer, Denis and Gale, Julian D. and Cölfen, Helmut},
	urldate = {2025-11-05},
	date = {2022},
    year = {2022},
	langid = {english},
}

@article{du_non-classical_2024,
	title = {Non-classical crystallization in soft and organic materials},
	volume = {9},
	issn = {2058-8437},
	url = {https://www.nature.com/articles/s41578-023-00637-y},
	doi = {10.1038/s41578-023-00637-y},
	pages = {229--248},
	number = {4},
	journal = {Nature Reviews Materials},
	shortjournal = {Nat Rev Mater},
	author = {Du, Jingshan S. and Bae, Yuna and De Yoreo, James J.},
	urldate = {2025-11-05},
	date = {2024-02-06},
    year = {2024},
	langid = {english},
}

@article{wu_nucleation_2022,
	title = {Nucleation and growth in solution synthesis of nanostructures – From fundamentals to advanced applications},
	volume = {123},
	issn = {00796425},
	url = {https://linkinghub.elsevier.com/retrieve/pii/S0079642521000451},
	doi = {10.1016/j.pmatsci.2021.100821},
	pages = {100821},
	journal = {Progress in Materials Science},
	shortjournal = {Progress in Materials Science},
	author = {Wu, Ke-Jun and Tse, Edmund C.M. and Shang, Congxiao and Guo, Zhengxiao},
	urldate = {2025-11-05},
	date = {2022-01},
    year = {2022},
	langid = {english},
}

@article{finney_molecular_2024,
	title = {Molecular simulation approaches to study crystal nucleation from solutions: Theoretical considerations and computational challenges},
	volume = {14},
	issn = {1759-0884},
	url = {https://onlinelibrary.wiley.com/doi/abs/10.1002/wcms.1697},
	doi = {10.1002/wcms.1697},
	shorttitle = {Molecular simulation approaches to study crystal nucleation from solutions},
	pages = {e1697},
	number = {1},
	journal = {{WIREs} Computational Molecular Science},
	author = {Finney, Aaron R. and Salvalaglio, Matteo},
	urldate = {2025-11-05},
	date = {2024},
    year = {2024},
	langid = {english},
}

@article{fu_recent_2022,
	title = {Recent Advances in Nonclassical Crystallization: Fundamentals, Applications, and Challenges},
	volume = {22},
	rights = {https://doi.org/10.15223/policy-029},
	issn = {1528-7483, 1528-7505},
	url = {https://pubs.acs.org/doi/10.1021/acs.cgd.1c01084},
	doi = {10.1021/acs.cgd.1c01084},
	shorttitle = {Recent Advances in Nonclassical Crystallization},
	pages = {1476--1499},
	number = {2},
	journal = {Crystal Growth \& Design},
	shortjournal = {Crystal Growth \& Design},
	author = {Fu, Haoyang and Gao, Xing and Zhang, Xin and Ling, Lan},
	urldate = {2025-11-05},
	date = {2022-02-02},
    year = {2022},
	langid = {english},
}

@article{jia_new_2024,
	title = {A New Perspective on Crystal Nucleation: A Classical View on Non-Classical Nucleation},
	volume = {24},
	rights = {https://doi.org/10.15223/policy-029},
	issn = {1528-7483, 1528-7505},
	url = {https://pubs.acs.org/doi/10.1021/acs.cgd.3c01078},
	doi = {10.1021/acs.cgd.3c01078},
	shorttitle = {A New Perspective on Crystal Nucleation},
	pages = {601--612},
	number = {1},
	journal = {Crystal Growth \& Design},
	shortjournal = {Crystal Growth \& Design},
	author = {Jia, Caiyun and Xiao, Anni and Zhao, Jiang and Wang, Pujun and Fang, Xiaoxia and Zhang, Haijun and Guan, Baohong},
	urldate = {2025-11-05},
	date = {2024-01-03},
    year = {2024},
	langid = {english},
}

@article{virtanen_scipy_2020,
	title = {{SciPy} 1.0: fundamental algorithms for scientific computing in Python},
	volume = {17},
	rights = {2020 The Author(s)},
	issn = {1548-7105},
	url = {https://www.nature.com/articles/s41592-019-0686-2},
	doi = {10.1038/s41592-019-0686-2},
	shorttitle = {{SciPy} 1.0},
	pages = {261--272},
	number = {3},
	journal = {Nature Methods},
	shortjournal = {Nat Methods},
	author = {Virtanen, Pauli and Gommers, Ralf and Oliphant, Travis E. and Haberland, Matt and Reddy, Tyler and Cournapeau, David and Burovski, Evgeni and Peterson, Pearu and Weckesser, Warren and Bright, Jonathan and van der Walt, Stéfan J. and Brett, Matthew and Wilson, Joshua and Millman, K. Jarrod and Mayorov, Nikolay and Nelson, Andrew R. J. and Jones, Eric and Kern, Robert and Larson, Eric and Carey, C. J. and Polat, İlhan and Feng, Yu and Moore, Eric W. and {VanderPlas}, Jake and Laxalde, Denis and Perktold, Josef and Cimrman, Robert and Henriksen, Ian and Quintero, E. A. and Harris, Charles R. and Archibald, Anne M. and Ribeiro, Antônio H. and Pedregosa, Fabian and van Mulbregt, Paul},
	urldate = {2024-05-22},
	date = {2020-03},
    year = {2020},
	langid = {english},
}

@article{ayyer_dragonfly_2016,
	title = {\textit{Dragonfly} : an implementation of the expand–maximize–compress algorithm for single-particle imaging},
	volume = {49},
	issn = {1600-5767},
	url = {https://journals.iucr.org/paper?S1600576716008165},
	doi = {10.1107/S1600576716008165},
	shorttitle = {\textit{Dragonfly}},
	pages = {1320--1335},
	number = {4},
	journal = {Journal of Applied Crystallography},
	shortjournal = {J Appl Crystallogr},
	author = {Ayyer, Kartik and Lan, Ti-Yen and Elser, Veit and Loh, N. Duane},
	urldate = {2025-10-22},
	date = {2016-08-01},
    year = {2016},
	langid = {english},
}

@article{ilett_studying_2024,
	title = {Studying crystallisation processes using electron microscopy: The importance of sample preparation},
	volume = {295},
	issn = {1365-2818},
	url = {https://onlinelibrary.wiley.com/doi/abs/10.1111/jmi.13300},
	doi = {10.1111/jmi.13300},
	shorttitle = {Studying crystallisation processes using electron microscopy},
	pages = {243--256},
	number = {3},
	journal = {Journal of Microscopy},
	author = {Ilett, Martha and Afzali, Maryam and Abdulkarim, Bilal and Aslam, Zabeada and Foster, Stephanie and Burgos-Ruiz, Miguel and Kim, Yi-Yeoun and Meldrum, Fiona C. and Drummond-Brydson, Rik M.},
	urldate = {2025-11-10},
	date = {2024},
    year = {2024},
	langid = {english},
}

@article{colfen_analytical_2023,
	title = {Analytical ultracentrifugation in colloid and polymer science: new possibilities and perspectives after 100 years},
	volume = {301},
	issn = {0303-402X, 1435-1536},
	url = {https://link.springer.com/10.1007/s00396-023-05130-0},
	doi = {10.1007/s00396-023-05130-0},
	shorttitle = {Analytical ultracentrifugation in colloid and polymer science},
	pages = {821--849},
	number = {7},
	journal = {Colloid and Polymer Science},
	shortjournal = {Colloid Polym Sci},
	author = {Cölfen, Helmut},
	urldate = {2025-11-10},
	date = {2023-07},
    year = {2023},
	langid = {english},
}

@article{wang_research_2022,
	title = {Research on Mesoscale Nucleation and Growth Processes in Solution Crystallization: A Review},
	volume = {12},
	rights = {http://creativecommons.org/licenses/by/3.0/},
	issn = {2073-4352},
	url = {https://www.mdpi.com/2073-4352/12/9/1234},
	doi = {10.3390/cryst12091234},
	pages = {1234},
	number = {9},
	journal = {Crystals},
	author = {Wang, Xiaowei and Li, Kangli and Qin, Xueyou and Li, Mingxuan and Liu, Yanbo and An, Yanlong and Yang, Wulong and Chen, Mingyang and Ouyang, Jinbo and Gong, Junbo},
	urldate = {2025-11-05},
	date = {2022-09},
    year = {2022},
	langid = {english},
}

@article{bols_situ_2024,
	title = {\textit{In Situ} {UV}–Vis–{NIR} Absorption Spectroscopy and Catalysis},
	volume = {124},
	rights = {https://doi.org/10.15223/policy-029},
	issn = {0009-2665, 1520-6890},
	url = {https://pubs.acs.org/doi/10.1021/acs.chemrev.3c00602},
	doi = {10.1021/acs.chemrev.3c00602},
	pages = {2352--2418},
	number = {5},
	journal = {Chemical Reviews},
	shortjournal = {Chem. Rev.},
	author = {Bols, Max L. and Ma, Jing and Rammal, Fatima and Plessers, Dieter and Wu, Xuejiao and Navarro-Jaén, Sara and Heyer, Alexander J. and Sels, Bert F. and Solomon, Edward I. and Schoonheydt, Robert A.},
	urldate = {2025-11-11},
	date = {2024-03-13},
    year = {2024},
	langid = {english},
}

@article{whitehead_particle_2021,
	title = {Particle formation mechanisms supported by \textit{in situ} synchrotron {XAFS} and {SAXS} studies: a review of metal, metal-oxide, semiconductor and selected other nanoparticle formation reactions},
	volume = {2},
	issn = {2633-5409},
	url = {https://xlink.rsc.org/?DOI=D1MA00222H},
	doi = {10.1039/D1MA00222H},
	shorttitle = {Particle formation mechanisms supported by \textit{in situ} synchrotron {XAFS} and {SAXS} studies},
	pages = {6532--6568},
	number = {20},
	journal = {Materials Advances},
	shortjournal = {Mater. Adv.},
	author = {Whitehead, Christopher B. and Finke, Richard G.},
	urldate = {2025-11-11},
	date = {2021},
    year = {2021},
	langid = {english},
}

@article{chapman_femtosecond_2011,
	title = {Femtosecond X-ray protein nanocrystallography},
	volume = {470},
	issn = {0028-0836, 1476-4687},
	url = {http://www.nature.com/articles/nature09750},
	doi = {10.1038/nature09750},
	pages = {73--77},
	number = {7332},
	journal = {Nature},
	shortjournal = {Nature},
	author = {Chapman, Henry N. and Fromme, Petra and Barty, Anton and White, Thomas A. and Kirian, Richard A. and Aquila, Andrew and Hunter, Mark S. and Schulz, Joachim and {DePonte}, Daniel P. and Weierstall, Uwe and Doak, R. Bruce and Maia, Filipe R. N. C. and Martin, Andrew V. and Schlichting, Ilme and Lomb, Lukas and Coppola, Nicola and Shoeman, Robert L. and Epp, Sascha W. and Hartmann, Robert and Rolles, Daniel and Rudenko, Artem and Foucar, Lutz and Kimmel, Nils and Weidenspointner, Georg and Holl, Peter and Liang, Mengning and Barthelmess, Miriam and Caleman, Carl and Boutet, Sébastien and Bogan, Michael J. and Krzywinski, Jacek and Bostedt, Christoph and Bajt, Saša and Gumprecht, Lars and Rudek, Benedikt and Erk, Benjamin and Schmidt, Carlo and Hömke, André and Reich, Christian and Pietschner, Daniel and Strüder, Lothar and Hauser, Günter and Gorke, Hubert and Ullrich, Joachim and Herrmann, Sven and Schaller, Gerhard and Schopper, Florian and Soltau, Heike and Kühnel, Kai-Uwe and Messerschmidt, Marc and Bozek, John D. and Hau-Riege, Stefan P. and Frank, Matthias and Hampton, Christina Y. and Sierra, Raymond G. and Starodub, Dmitri and Williams, Garth J. and Hajdu, Janos and Timneanu, Nicusor and Seibert, M. Marvin and Andreasson, Jakob and Rocker, Andrea and Jönsson, Olof and Svenda, Martin and Stern, Stephan and Nass, Karol and Andritschke, Robert and Schröter, Claus-Dieter and Krasniqi, Faton and Bott, Mario and Schmidt, Kevin E. and Wang, Xiaoyu and Grotjohann, Ingo and Holton, James M. and Barends, Thomas R. M. and Neutze, Richard and Marchesini, Stefano and Fromme, Raimund and Schorb, Sebastian and Rupp, Daniela and Adolph, Marcus and Gorkhover, Tais and Andersson, Inger and Hirsemann, Helmut and Potdevin, Guillaume and Graafsma, Heinz and Nilsson, Björn and Spence, John C. H.},
	urldate = {2021-01-11},
	date = {2011-02},
    year = {2011},
	langid = {english},
}

@article{vreshch_monomeric_2010,
	title = {Monomeric Square-Planar Cobalt({II}) Acetylacetonate: Mystery or Mistake?},
	volume = {49},
	issn = {0020-1669},
	url = {https://doi.org/10.1021/ic100963r},
	doi = {10.1021/ic100963r},
	shorttitle = {Monomeric Square-Planar Cobalt({II}) Acetylacetonate},
	pages = {8430--8434},
	number = {18},
	journal = {Inorganic Chemistry},
	shortjournal = {Inorg. Chem.},
	author = {Vreshch, Volodimir D. and Yang, Jen-Hsien and Zhang, Haitao and Filatov, Alexander S. and Dikarev, Evgeny V.},
	urldate = {2025-11-12},
	date = {2010-09-20},
    year = {2010},
}

@article{chrzanowski_-tris24-pentanedionato-2oocobaltiii_2007,
	title = {{$\alpha$}-Tris(2,4-pentanedionato-{$\kappa$}${\sp 2}${\it O},{\it O}{$^\prime$})cobalt(III) at 240, 210, 180, 150 and 110K},
	volume = {63},
	issn = {0108-2701},
	url = {https://journals.iucr.org/c/issues/2007/07/00/gz3091/},
	doi = {10.1107/S0108270107022950},
	pages = {m283--m288},
	number = {7},
	journal = {Acta Crystallographica Section C: Crystal Structure Communications},
	shortjournal = {Acta Cryst C},
	author = {Chrzanowski, L. S. von and Lutz, M. and Spek, A. L.},
	urldate = {2025-11-12},
	date = {2007-07-15},
    year = {2007},
	langid = {english},
}

@article{v_zimmermann_p211_2025,
	title = {P21.1 at {PETRA} {III} – a high-energy X-ray diffraction beamline for physics and chemistry},
	volume = {32},
	rights = {https://creativecommons.org/licenses/by/4.0/},
	issn = {1600-5775},
	url = {https://journals.iucr.org/s/issues/2025/03/00/yi5172/},
	doi = {10.1107/S1600577525002826},
	pages = {802--814},
	number = {3},
	journal = {Journal of Synchrotron Radiation},
	shortjournal = {J Synchrotron Rad},
	author = {v. Zimmermann, M. and Ivashko, O. and Igoa Saldaña, F. and Liu, J. and Glaevecke, P. and Gutowski, O. and Nowak, R. and Köhler, K. and Winkler, B. and Schöps, A. and Schulte-Schrepping, H. and Dippel, A.-C.},
	urldate = {2025-05-05},
	date = {2025-05-01},
    year = {2025},
	langid = {english},
}

@article{kieffer_pyfai_2013,
	title = {{PyFAI}: a Python library for high performance azimuthal integration on {GPU}},
	volume = {28},
	issn = {0885-7156, 1945-7413},
	url = {https://www.cambridge.org/core/journals/powder-diffraction/article/pyfai-a-python-library-for-high-performance-azimuthal-integration-on-gpu/907C9E3D59DED712C55629CEB988C27B},
	doi = {10.1017/S0885715613000924},
	shorttitle = {{PyFAI}},
	pages = {S339--S350},
	issue = {S2},
	journal = {Powder Diffraction},
	author = {Kieffer, J. and Wright, J. P.},
	urldate = {2024-06-04},
	date = {2013-09},
    year = {2013},
	langid = {english},
}

@article{juhas_pdfgetx3_2013,
	title = {\textit{{PDFgetX}3} : a rapid and highly automatable program for processing powder diffraction data into total scattering pair distribution functions},
	volume = {46},
	issn = {0021-8898},
	url = {http://scripts.iucr.org/cgi-bin/paper?S0021889813005190},
	doi = {10.1107/S0021889813005190},
	shorttitle = {\textit{{PDFgetX}3}},
	pages = {560--566},
	number = {2},
	journal = {Journal of Applied Crystallography},
	shortjournal = {J Appl Crystallogr},
	author = {Juhás, P. and Davis, T. and Farrow, C. L. and Billinge, S. J. L.},
	urldate = {2020-09-14},
	date = {2013-04-01},
    year = {2013},
	langid = {english},
}

@article{juhas_complex_2015,
	title = {Complex modeling: a strategy and software program for combining multiple information sources to solve ill posed structure and nanostructure inverse problems},
	volume = {71},
	issn = {2053-2733},
	url = {https://journals.iucr.org/a/issues/2015/06/00/ae5008/},
	doi = {10.1107/S2053273315014473},
	shorttitle = {Complex modeling},
	pages = {562--568},
	number = {6},
	journal = {Acta Crystallographica Section A: Foundations and Advances},
	shortjournal = {Acta Cryst A},
	author = {Juhás, P. and Farrow, C. and Yang, X. and Knox, K. and Billinge, S.},
	urldate = {2025-06-04},
	date = {2015-11-01},
    year = {2015},
	langid = {english},
}

@article{jia_dynamic_2023,
	title = {Dynamic Light Scattering: A Powerful Tool for In Situ Nanoparticle Sizing},
	volume = {7},
	rights = {http://creativecommons.org/licenses/by/3.0/},
	issn = {2504-5377},
	url = {https://www.mdpi.com/2504-5377/7/1/15},
	doi = {10.3390/colloids7010015},
	shorttitle = {Dynamic Light Scattering},
	pages = {15},
	number = {1},
	journal = {Colloids and Interfaces},
	author = {Jia, Zixian and Li, Jiantao and Gao, Lin and Yang, Dezheng and Kanaev, Andrei},
	urldate = {2025-11-14},
	date = {2023-03},
    year = {2023},
	langid = {english},
}

@article{mancuso_single_2019,
  title = {The single particles, clusters and biomolecules and serial femtosecond crystallography instrument of the European XFEL: initial installation},
  author = {Mancuso, Adrian P. and Aquila, Andrew and Batchelor, Lewis and Bean, Richard J. and Bielecki, Johan and Borchers, Gannon and Doerner, Katerina and Giewekemeyer, Klaus and Graceffa, Rita and Kelsey, Oliver D. and Kim, Yoonhee and Kirkwood, Henry J. and Legrand, Alexis and Letrun, Romain and Manning, Bradley and Lopez Morillo, Luis and Messerschmidt, Marc and Mills, Grant and Raabe, Steffen and Reimers, Nadja and Round, Adam and Sato, Tokushi and Schulz, Joachim and Signe Takem, Cedric and Sikorski, Marcin and Stern, Stephan and Thute, Prasad and Vagovi{\v{c}}, Patrik and Weinhausen, Britta and Tschentscher, Thomas},
  journal = {Synchrotron Radiation},
  volume = {26},
  number = {3},
  pages = {660--676},
  year = {2019},
}

@article{bielecki_electrospray_2019,
  title={Electrospray sample injection for single-particle imaging with x-ray lasers},
  author={Johan Bielecki  and Max F. Hantke  and Benedikt J. Daurer  and Hemanth K. N. Reddy  and Dirk Hasse  and Daniel S. D. Larsson  and Laura H. Gunn  and Martin Svenda  and Anna Munke  and Jonas A. Sellberg  and Leonie Flueckiger  and Alberto Pietrini  and Carl Nettelblad  and Ida Lundholm  and Gunilla Carlsson  and Kenta Okamoto  and Nicusor Timneanu  and Daniel Westphal  and Olena Kulyk  and Akifumi Higashiura  and Gijs van der Schot  and Ne-Te Duane Loh  and Taylor E. Wysong  and Christoph Bostedt  and Tais Gorkhover  and Bianca Iwan  and M. Marvin Seibert  and Timur Osipov  and Peter Walter  and Philip Hart  and Maximilian Bucher  and Anatoli Ulmer  and Dipanwita Ray  and Gabriella Carini  and Ken R. Ferguson  and Inger Andersson  and Jakob Andreasson  and Janos Hajdu  and Filipe R. N. C. Maia },
  journal={Science Advances},
  volume={5},
  number={5},
  pages={eaav8801},
  year={2019},
  publisher={American Association for the Advancement of Science}
}

@article{allahgholi_megapixels_2019,
  title={Megapixels@ Megahertz--The AGIPD high-speed cameras for the European XFEL},
  author={Aschkan Allahgholi and Julian Becker and Annette Delfs and Roberto Dinapoli and Peter Göttlicher and Heinz Graafsma and Dominic Greiffenberg and Helmut Hirsemann and Stefanie Jack and Alexander Klyuev and Hans Krüger and Manuela Kuhn and Torsten Laurus and Alessandro Marras and Davide Mezza and Aldo Mozzanica and Jennifer Poehlsen and Ofir {Shefer Shalev} and Igor Sheviakov and Bernd Schmitt and Jörn Schwandt and Xintian Shi and Sergej Smoljanin and Ulrich Trunk and Jiaguo Zhang and Manfred Zimmer},
  journal={Nuclear Instruments and Methods in Physics Research Section A: Accelerators, Spectrometers, Detectors and Associated Equipment},
  volume={942},
  pages={162324},
  year={2019},
  publisher={Elsevier}
}

@article{loh_reconstruction_2009,
  title={Reconstruction algorithm for single-particle diffraction imaging experiments},
  author={Loh, Ne-Te Duane and Elser, Veit},
  journal={Physical Review E—Statistical, Nonlinear, and Soft Matter Physics},
  volume={80},
  number={2},
  pages={026705},
  year={2009},
  publisher={APS}
}

\end{document}